%% file: planck_vla.tex
\documentclass[preprint2]{aastex}
\pdfoutput=1
\usepackage{amsmath,amsfonts,amssymb}
\usepackage{txfonts}
\usepackage{graphicx}
\usepackage[breaklinks, colorlinks, citecolor=blue]{hyperref}
\usepackage{fixltx2e}
\usepackage{natbib}
\bibpunct{(}{)}{;}{a}{}{,} 

\input Planck.tex

\def\Plancks{\textit{Planck}'s}

\shorttitle{Absolute Calibration of the Radio Astronomy Flux Density Scale}
\shortauthors{Partridge et al.}

\begin{document}

\title{Absolute Calibration of the Radio Astronomy Flux Density Scale at 22 to 43\,GHz Using \Planck\ }
\author{B. Partridge\altaffilmark{1}, M. L\'opez-Caniego\altaffilmark{2,3}, R. A. Perley \altaffilmark{4},  J. Stevens\altaffilmark{5},  B. J. Butler\altaffilmark{4},  G. Rocha\altaffilmark{6,7}, B. Walter\altaffilmark{1},  A. Zacchei\altaffilmark{8} }
\affil{\altaffilmark{1} Haverford College Astronomy Department, 370 Lancaster Avenue, Haverford, Pennsylvania, U.S.A. }
\affil{\altaffilmark{2} European Space Agency, ESAC, Camino bajo del Castillo, s/n, Urbanizaci\'{o}n Villafranca del Castillo, Villanueva de la Ca\~{n}ada, Madrid, Spain}
\affil{\altaffilmark{3} Instituto de F\'{\i}sica de Cantabria (CSIC-Universidad de Cantabria), Avda. de los Castros s/n, Santander, Spain}
\affil{\altaffilmark{4} National Radio Astronomy Observatory, P.O. Box O, Socorro, NM, 87801, USA}
\affil{\altaffilmark{5} CSIRO Astronomy and Space Science, Paul Wild Observatory, 1828 Yarrie Lake Road, Narrabri, NSW 2390, Australia}
\affil{\altaffilmark{6} Jet Propulsion Laboratory, California Institute of Technology, 4800 Oak Grove Drive, Pasadena, California, U.S.A.}
\affil{\altaffilmark{7} California Institute of Technology, Pasadena, California, U.S.A.}
\affil{\altaffilmark{8} INAF - Osservatorio Astronomico di Trieste, Via G.B. Tiepolo 11, Trieste, Italy}

\begin{abstract}
The \Planck\ mission detected thousands of extragalactic radio sources at frequencies from 28 to 857\,GHz.  \Plancks\ calibration is absolute (in the sense that it is based on the satellite's annual motion around the Sun and the temperature of the cosmic microwave background), and its beams are well-characterized at sub-percent levels.  Thus \Plancks\ flux density measurements of compact sources are absolute in the same sense.  We have made coordinated VLA and ATCA observations of 65 strong, unresolved \Planck\ sources in order to transfer \Plancks\ calibration to ground-based instruments at 22, 28, and 43\,GHz.  The results are compared to microwave flux density scales currently based on planetary observations.  Despite the scatter introduced by the variability of many of the sources, the flux density scales are determined to $1-2\%$ accuracy.  At 28\,GHz, the flux density scale used by the VLA runs $2-3\% \pm 1.0\%$ below \Planck\ values with an uncertainty of $\pm 1.0\%$; at 43\,GHz, the discrepancy increases to $5-6\% \pm 1.4\%$ for both ATCA and the VLA.
\end{abstract}


\maketitle

\section{Introduction}\label{sec:INTRO}
Calibration of the flux density scale used by radio astronomers was for many years based on observations of a set of strong radio sources made with scaled horns or other instruments having well-determined optical properties \citep{Baars77}.   More recently, flux density scales have been revised by \cite{Perley13} in the 1--50\,GHz frequency range, based on extensive observations of Mars made at the Karl G. Jansky Very Large Array (VLA) operated by NRAO.\footnote{The National Radio Astronomy Observatory is a facility of the National Science Foundation operated under cooperative agreement by Associated Universities, Inc.}  A similar calibration at 30\,GHz pinned to observations of Jupiter is presented by \cite{Hafez08}.  This calibration method depends on accurate knowledge of the planet's surface temperature and its variation over time.  Perley \& Butler used planetary temperatures adjusted to fit extensive observations of Mars by the \textit{WMAP} satellite \citep{Weiland11}.  The \textit{WMAP} measurements are important because the \textit{WMAP} calibration is absolute, since it is determined from the dipole signal induced in the 2.7255\,K cosmic background radiation (CMB) by the satellite's yearly motion around the Sun \citep{Hinshaw09,Fixen09}.

\subsection{Planck-based calibration}
The European Space Agency's \Planck\footnote{\Planck\ (\url{http://www.esa.int/Planck}) is a project of the European Space Agency (ESA) with instruments provided by two scientific consortia funded by ESA member states (in particular the lead countries France and Italy), with contributions from NASA (USA) and telescope reflectors provided by a collaboration between ESA and a scientific consortium led and funded by Denmark.}  mission, like \textit{WMAP}, is calibrated absolutely from the CMB dipole \citep{planck2014-a01}.  \Plancks\ higher resolution and greater sensitivity permit a more direct method of transferring its absolute calibration to ground-based radio telescopes.  This paper describes the results obtained by this method. It is based on approximately simultaneous (explained below) observations of many strong radio sources using \Planck\, and the more sensitive VLA and Australia Telescope Compact Array (ATCA).\footnote{ATCA (http://www.narrabri.atnf.csiro.au) is funded by the Commonwealth of Australia for operation as a National Facility managed by CSIRO.} The results reported here are based on a more thorough analysis than preliminary results reported in a recent \Planck\ paper \citep{planck2014-a35}.

Several of the standard sources calibrated by \cite{Perley13}, such as 3C\,48 and 3C\,286, were detected by \Planck\, but at low significance; hence we made the choice to observe a set of stronger calibration sources.  We also observed scores of sources rather than concentrating on a few with high flux densities as a further control over the variability of radio sources at high frequencies.  Finally, linear polarization was measured for each source.

\begin{figure*} 
\begin{center}
\includegraphics[width=0.95\textwidth]{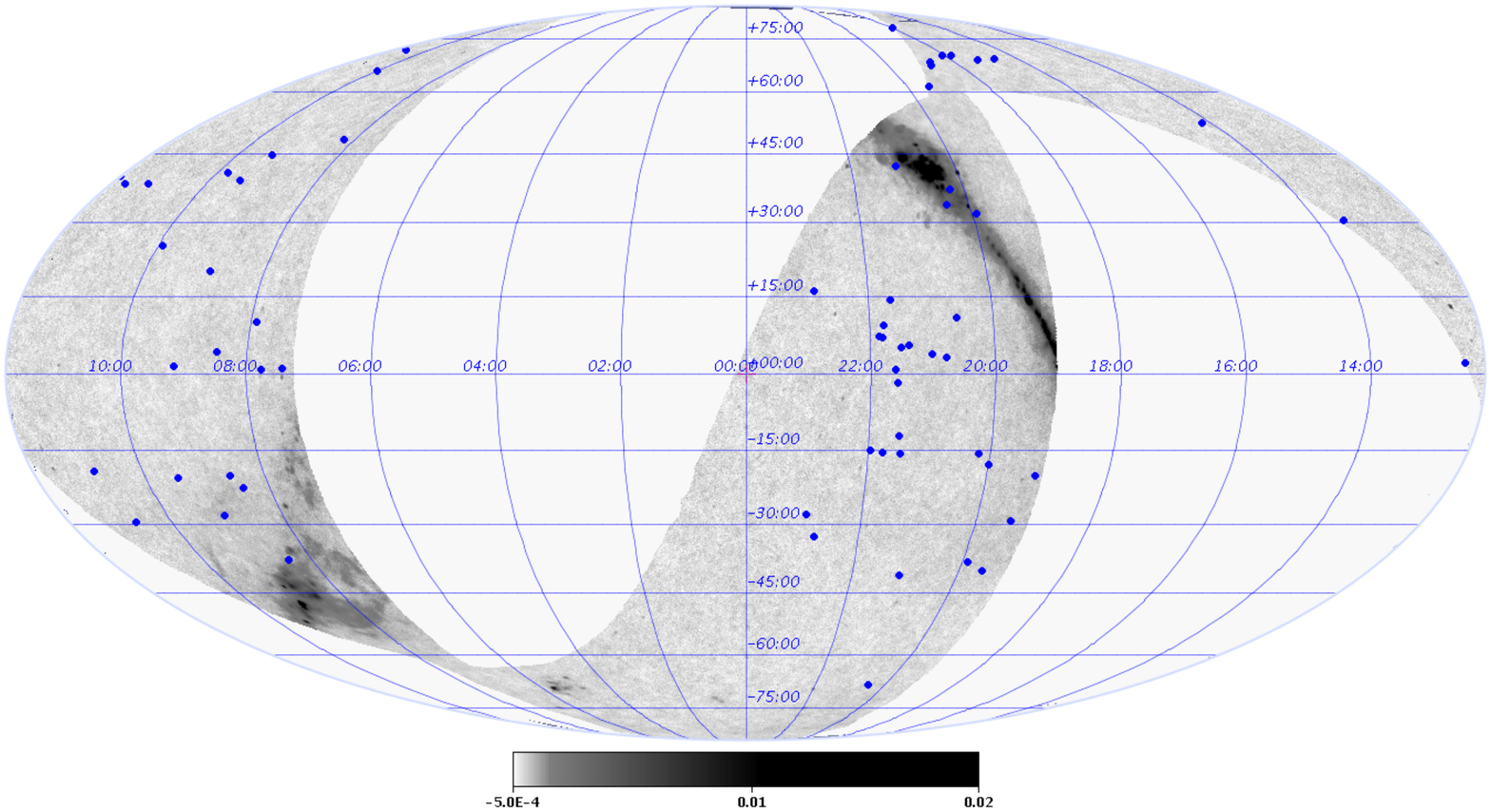}
\caption{Sky distribution of the sources on the 30\,GHz special \textit{Planck} map that covers the observing period 2013 April 1 to June 30. The figure is a full-sky Mollweide projection in equatorial coordinates. The blank unobserved pixels are shown in white and the units of the colorbar are kelvins.}
\label{fig:sky_dist}
\end{center}
\end{figure*}

The \Planck\ scan strategy \citep{planck2011-1.1} is fixed.  Thus the date at which a source at particular celestial coordinates was observed can be found, for instance, by the POFF tool \citep{Massardi10}.  This information allowed us to coordinate the \Planck\ and ground-based observations.  The VLA is dynamically scheduled, so we did not know in advance the exact dates of these ground-based observations.  We therefore selected sources that \Planck\ was scheduled to scan sometime in the three-month period of 2013 April--June.  The list included sources near the ecliptic poles, regions of the sky covered nearly continuously by \textit{Planck}, as well as low-declination sources visible to the ground-based instruments in both hemispheres.  We also included some fainter sources to allow us to increase the flux density range and to test the linearity of the flux density scales used at the VLA and ATCA; the direct VLA--ATCA comparison especially for frequencies lower than \Plancks\ 28\,GHz band,  will be treated in a separate paper \citep{Stevens15}.  The locations of these sources, and the area of the sky scanned by \Planck\ in the period 2013 April 1 to 2013 June 30, are shown in Figure~\ref{fig:sky_dist}.  

VLA observations were made at two epochs, roughly 4 weeks apart, to provide some information on the possible variability of the sources we used.  The ATCA observations spanned a period of approximately two weeks in 2013 April.  The issue of source variability is discussed further in Section~\ref{sec:source_var}.
 
\subsection{Outline}  
In Sections~\ref{sec:JVLAObs}, \ref{sec:ATCAObs}, and \ref{sec:PLANCK}, we discuss the observations made with the VLA, ATCA, and \textit{Planck}, respectively, and the methods used to determine flux densities from each instrument.  The comparison of ground-based and satellite flux densities at 22, 28, and 43\,GHz is made in Section~\ref{sec:COMP}.  Polarization measurements are very briefly discussed in Section~\ref{sec:POL}, and we summarize and discuss the results in Section~\ref{sec:conclusions}.  

This paper addresses the consistency of the flux-density scales used at the VLA and ATCA only at frequencies above $\sim$20\,GHz, where the \Planck\ data can be employed.  A separate paper \citep{Stevens15} treats the comparison of measurements at the two ground-based instruments made at lower frequencies.

\section{VLA Observations and Data Reduction }\label{sec:JVLAObs}
The VLA observations were taken in two sessions, 30\,h on 2013 May 3--4, while the array was in the most compact `D' configuration, and 18\,h on 2013 May 30, while the array was being reconfigured between `D' and `C' configurations. These observations were part of a regular observatory maintenance program; the data are used to check system performance, and to determine various system parameters.  Observations were made in eight VLA frequency bands, using the 8-bit samplers, which limit the total bandwidth in each frequency band to 2048\,MHz.  The data from each observing band employed in this paper are organized into 16 sub-bands, each of 128\,MHz.  The frequencies spanned are given in Table~\ref{tab:JVLA}.  Only the data at frequencies above 20\,GHz are used here; the lower frequency observations are discussed in \cite{Stevens15}.  Because the observed sources are all fairly strong, there is no need to use the full bandwidth for the imaging; only a single 128\,MHz-wide sub-band is needed.  The center frequencies actually used to determine the flux density and polarization of the sources are given in Table~\ref{tab:JVLA}.  

\begin{table}
\caption{Frequencies employed at the VLA. \label{tab:JVLA}}
\smallskip
\begin{tabular}{lcc}
\tableline
\tableline
Band&Frequency span&Central frequencies\\ 
&(GHz)&(GHz)\\
\tableline
L & 0.985 -- 2.025 & 1.465, 1.865 \\
S & 1.989 -- 4.013 & 2.565, 3.565 \\
C & 4.309 -- 5.333 & 4.885 \\
 & 6.437 -- 7.461 & 6.885 \\
X & 8.243 -- 9.267 & 8.435 \\
 & 10.488 -- 11.512 & 11.064 \\
Ku & 14.133 -- 15.157 & 14.965 \\
 & 16.976 -- 18.000 & 17.422 \\
K & 21.618 -- 22.642 & 22.450 \\
& 25.388 -- 26.412 & 25.836 \\
Ka & 28.258 -- 29.282 & 28.450 \\
& 35.987 -- 37.011 & 36.435 \\
Q & 43.148 -- 43.256 & 43.340 \\
& 47.849 -- 48.873 & 48.425 \\
\tableline
\end{tabular}
\end{table}


The data were calibrated with the AIPS calibration package, using the regimen described by \cite{Perley13} as well as their revised flux density scale.  The calibration took account of atmospheric opacity.  Polarization calibration was established following the regimen described by \cite{Perley13b}, including the adjusted position angles for 3C\,286 described in that paper. 

Accurate flux density measurements using the VLA require corrections for three factors which change the antenna gains.  (1)  Changes in electronic gains are monitored by injecting at 10\,Hz a small amount of wideband noise into the receiver from a stable noise diode.  This contributed power is detected by the correlator. Variations in this detected power are proportional to changes in the receiver gain, and are used to correct the visibilities for these changes.  (2) Errors in antenna pointing are minimized by employing the referenced pointing technique, whereby the local antenna pointing offsets are determined using a nearby calibrator source, and the offsets are then employed for the target source.  For this set of observations, all sources were strong enough to employ the technique directly.  This reduced the typical pointing error from 10--20 arcsec to $\sim5$ arcsec.  (3) Finally, variations in antenna gain with elevation are measured by fitting a second-order polynomial in elevation to the measurements of sources of known flux density during the observations.  There are typically 30 observations of such sources, over the full range of elevation available (typically 10--80 degrees), permitting an estimate of the antenna gain as a function of elevation to better than 1\%.  At 28\,GHz, the typical change in power gain of the VLA antennas is $\sim10\%$.  At 43\,GHz, it can be as high as 30\%.

Following these corrections, the flux density scale is set by observations of 3C\,286, a source known to be stable over the past thirty years \citep{Perley13}.  At all frequencies above 10\,GHz, the limiting factor in determining accurate flux densities with the VLA are the residual pointing errors (see \citealt{Perley13} for a fuller discussion).

Flux density determinations were made through directly imaging the sources in Stokes parameters $I$, $Q$, and $U$, using well-established techniques for self-calibration and deconvolution (see \citealt{Perley13} for details).  The VLA resolution varied because of the the configuration employed as well as the different declinations, $\delta$, of the sources observed.  Typically, the resolution for the 'D' configuration data is given by $[60 \times 60 \csc(\delta-35^\circ)]/\nu$ arcsec, with $\nu$ in gigahertz.  For the second observing epoch, taken in a mixed configuration, the beam sizes are about half this size.  The total flux density for a source was determined by integration of the source strength over an area encompassing all visible emission.  This approach was taken since the \Planck\ beams were far larger than the VLA beams. While we also computed peak brightness for each source, we consistently used total flux densities when making comparisons with \Planck\ measurements.  We also examined results in the visibility data to check for extended emission.

Any significant difference between total and peak brightness is an indication that the source may be resolved.  Clear evidence of resolution was found for several sources such as J2107+4213 (NGC\,7027) and J0813+4812 (3C\,196).  In Section~\ref{sec:EffResol}, we discuss the effect of excluding evidently resolved sources.

The uncertainties in these VLA flux density measurements were determined from the scatter in the individual observations of each source, as each object was typically observed five times during the course of each run.  Potential systematic error in the flux density scale introduced by Perley and Butler arises from uncertainties in the model for emission by Mars (the average dielectric constant of the surface; the extent of and changes in the polar caps, etc.) and in the transfer of measurements from Mars to 3C\,286.  These issues are discussed more fully in \cite{Perley13}, section 9.  There, the estimated uncertainty in the flux density scale at 22\,GHz is given as $\sim2\%$, rising to $\sim3\%$ at 43\,GHz.

\section{ATCA Observations and Data Reduction} \label{sec:ATCAObs}

\subsection{Observations}
The Australia Telescope Compact Array (ATCA) observed the comparison sources between 2013 April 17 and 29, during the \Planck\ observations described above (but days to weeks earlier than the VLA observations). The Compact Array Broadband Backend (CABB) system installed on ATCA \citep{Wilson10} provides two simultaneous sub-bands, each with 2048 MHz of bandwidth. On each day a different pair of simultaneous frequencies was observed, as detailed in Table \ref{tab:ATCA}.

During each observing epoch, the sources were observed alternately for a few minutes at a time, so that each was observed over a range of hour angles and elevations. Since all of the sources had flux densities above 100\,mJy, integration time was not a primary concern, and no source was observed for more than 25\,min per epoch. Between scans on the target sources, a number of potential flux density calibrators were observed. These were the planets Venus, Mars, Uranus, and Neptune, and the gigahertz-peaked spectrum (GPS) source PKS\,1934$-$638; this latter source is regularly used as the primary flux density calibrator for the ATCA for frequencies below 25\,GHz.  To limit the effect of antenna gain changes as a function of elevation, the calibrator sources were observed over the same range of elevations as the program sources. For at least one measurement for each source at each epoch, the elevation of the calibrator source PKS 1934-638 matched the program sources to better than 2.65 degrees. The measured gains vary by less than $1\%$ over such an elevation range for frequencies below 30 GHz, and by approximately $1\%$ in the worst case for higher frequencies. We therefore conclude that our measurements are not biased by systematic elevation dependencies.  The system temperature of each antenna was constantly monitored by injecting noise at the receiver front-end, and variations in system temperature were compensated for by scaling the amplitudes in the correlator.

The ATCA was in its H214 configuration during the observations. Five of the six ATCA antennas were separated by between 82 and 247\,m. Observations were scheduled in this compact array to better match \Plancks\ resolution, to minimize possible resolution of complex sources, and to limit the effect of the atmosphere at the higher frequencies.

\begin{table*}
\caption{Frequencies employed at ATCA. \label{tab:ATCA}}
\smallskip
\begin{tabular}{lccc}
\tableline\tableline
Observation epochs&Central frequency 1&Central frequency 2&Band\\ 
(UTC)&(GHz)&(GHz)& \\ 
\tableline
2013 Apr 17, 22, 29    &   44.00	   &  48.50	   &  7 mm   \\
2013 Apr 24    &   5.50	   &  9.00	   &  4 cm   \\
2013 Apr 20, 24, 29    &   18.50	   &  23.00	   &  15 mm   \\
\tableline
\end{tabular}
\end{table*}

\subsection{Data Reduction}
All data reduction was done with the ATCA software reduction package Miriad \citep{Sault11}. At the very beginning of the data reduction process, corrections for atmospheric opacity changes were made using meteorological data recorded during the runs; the Miriad package uses the atmospheric models of \cite{Liebe85} to make these corrections. The gains derived through our calibration process displayed less than 3\% variation over the range of zenith opacities experienced during our observations, indicating that our opacity corrections were successful to at least that precision. Elevation dependence of the gains was also corrected for at this stage.

\subsubsection{Flux density model for PKS\,1934$-$638}
The Miriad software has a built-in model for the flux density of PKS\,1934$-$638 as a function of frequency, based on the models of \cite{Reynolds94} and \cite{Sault03}. In this paper, we use a new model for PKS 1934-638 which is derived by including measurements of its flux density at frequencies between 92 and 96 GHz. These high-frequency measurements were performed on 2012 August 12, with the ATCA in its H75 configuration. Only five of the six antennas were used due to receiver constraints.  Observations were made at night, in excellent, stable conditions. PKS 1934-638 was observed for a total of 63 minutes, and its flux density was measured by comparing it to that observed for the planet Uranus, using the \cite{dePater90} model. Measurements were made in two independent 2048 MHz bands, centered at 93\,GHz and 95\,GHz. No opacity corrections were required during data reduction because atmospheric opacity changes are compensated for during the observations through regular observations of an absorbing paddle. Elevation-dependent gain changes were corrected for using the model of \cite{Subrahmanyan02}. In these observations, PKS 1934-638 covered the elevation range 32 to 44 degrees, while Uranus was observed at approximately 55 degrees elevation.
The flux density of PKS 1934-638 was measured from the images made in each band, and assigned to the central frequency of that band. Both images had a synthesized beam size of 7.70 x 5.17 arcseconds, and the flux density was measured to be the same in each image, to within the uncertainties, at 0.11 $\pm$ 0.01 Jy. These flux densities, measured at such high frequency, are an excellent constraint on the flux density model for PKS 1934-638.
 
We chose to modify the \cite{Sault03} flux density model to incorporate these higher frequency flux density measurements, while assuming that the \cite{Sault03} model provides correct flux densities in the range 16 - 24 GHz. We make this assumption in order that flux densities referenced against our modified model will closely match those made against the \cite{Sault03} model, which has been in use for ATCA data reduction since 2003.

To do this, we first evaluated the \cite{Sault03} model to get a list of the Stokes I flux density of PKS 1934-638 every 128 MHz between 10 GHz and 24 GHz. We assume a very conservative uncertainty of 0.1 Jy for each of these flux density values (slightly less than $10\%$ uncertainty). The 93 GHz and 95 GHz measurements of PKS 1934-638, as listed above, are added to this list, and a first-order linear least-squares fit is made. The resulting flux density model fit is:
\begin{equation}
\log S = 5.8870 - 1.3763 \log\nu,
\end{equation}
where $S$ is the flux density in Jy, $\nu$ is the frequency in MHz, and the log is base-10. This paper aims in part at evaluating how closely this proposed model agrees with the absolute calibration provided by \Planck\, without making any other presumptions about the quality of the proposed model.
 
\subsubsection{Calibration}
For each observation, data from two independent 2048\,MHz bands were independently reduced.  Calibration began with the determination of the bandpass response and an initial estimate of the time-dependent gain variation of one of the stronger sources observed during that epoch. In the 4\,cm band (4--11\,GHz), PKS\,1934$-$638 can be used for this purpose, since it is quite strong ( $>2$ Jy) and its flux density model is known; thus, the bandpass determination should not need further correction.

At higher frequencies we used the strong VLBI calibrator B1921$-293$ to determine the bandpass response. Because Miriad does not have a model for the flux density of B1921$-293$,  a flat ($\alpha = 0$) spectrum is the default during the bandpass determination. This is not entirely correct, but assuming that the actual flux density model has a power-law slope at these frequencies (an accurate assumption as it turns out) then a later correction for this slope is straightforward. This correction was done by transferring the bandpass solution to the flux density calibrator and then adjusting the bandpass slope to match the expected flux density behavior.

This flux calibrated bandpass solution was applied to all the other sources observed on the same day. The time-dependent gains were determined for each of the other sources independently. 

Once each source was self-calibrated in this way, the visibility plane data for all sources were rescaled to the bandpass calibrator's flux density scale, and corrections were made for any slope variations introduced in the gain determination.

\subsubsection{Measurements}
Flux density measurements were made using Miriad from the vector-averaged spectra of each source. A first or second order linear least-squares fit was made to the observed Stokes $I$ flux densities as a function of frequency over both of the simultaneously observed channels. The fit that best describes the spectra was used: this was determined by computing the rms of the residual amplitudes after the fitted model is subtracted from the spectra. 

The ATCA did not observe the exact same bands as the VLA, so the fitted ATCA flux density models were evaluated at the VLA frequency that lay closest.  The ATCA frequencies relevant for this paper are 17.422\,GHz and 22.450\,GHz in the 15\,mm band, and 43.340\,GHz and 48.425\,GHz in the 7\,mm band.
	
The Stokes $I$ flux densities were the direct result of this process.  The uncertainty for the Stokes $I$ measurements is the rms scatter of the spectral amplitudes around the model fit. 

\subsubsection{Measurement Uncertainty}
Since many of the comparison sources were observed in more than one of the seven epochs, we can look at the consistency of the measurements to get an estimate of the accuracy of our measurements.  This estimate will include any inconsistencies introduced by unrecognized variability of the sources.

In the 15\,mm band we observed eight sources in two separate epochs, and in the 7\,mm band we observed ten sources in two or more epochs, with seven observed in three epochs.

The observations of the eight multiply-observed sources in the 15\,mm band show that on average, the Stokes $I$ flux densities of these sources vary by 1.3\% between the epochs, and by no more than 3.7\%. In the 7\,mm band, the average Stokes $I$ variance is 5\%, and the maximum variance is 10.7\%. Since we cannot be certain that any of the sources varied intrinsically over the seven epochs, we have to assume conservatively that the actual measurement uncertainty could be as large as 3.7\% in the 15\,mm band and 10.7\% in the 7\,mm band.  

\section{\Planck\  Measurements}\label{sec:PLANCK}
The \Planck\ satellite \citep{tauber2010a,planck2013-p01} was launched on 2009 May 14, and scanned the sky stably and continuously from 2009 August 12 to 2013 October 23. \Planck\ carried a scientific payload consisting of an array of 74 detectors sensitive to a range of frequencies between 25 and 1000\,GHz, which scanned the sky simultaneously and continuously with an angular resolution varying between 30\,arcmin at the lowest frequencies and 5\,arcmin at the highest. The array is arranged into two instruments. The detectors of the Low Frequency Instrument \citep{Bersanelli10,planck2011-1.4} are pseudo-correlation radiometers, covering three bands centered at 28.4, 44.1, and 70.4\,GHz. The detectors of the High Frequency Instrument \citep{planck2011-1.5} are bolometers, covering six bands centered at 100, 143, 217, 353, 545, and 857\,GHz. The design of \Planck\ allows it to image the whole sky twice per year, with a combination of sensitivity, angular resolution and frequency coverage never before achieved.

For the results discussed here, flux densities at \Plancks\ three lowest frequencies were derived from special maps including only data taken during the period 2013 April 1 to 2013 June 30.  These maps were constructed by the LFI Data Processing Center (DPC) in Trieste (Italy).  Flux densities were derived using a non-blind approach at the position of each VLA or ATCA source using the Mexican Hat Wavelet 2 (MHW2) algorithm \citep{gnuevo06,caniego06}.  This algorithm preserves the amplitudes of compact sources while greatly reducing the effects of large scale structure (such as Galactic foregrounds, a random background of faint sources or fluctuations in the CMB) as well as small scale fluctuations (such as instrument noise). Further details about the implementation of the MHW2 are given in \cite{planck2014-a35}.

For all but the strongest sources (or those in confused regions at low Galactic latitude), flux density uncertainties for individual sources were $\sim$0.15\,Jy at 28\,GHz and $\sim$0.26\,Jy at 44\,GHz.  

The overall calibration uncertainty for the \Planck\ LFI instrument at 30, 44, and 70\,GHz was $0.35\%$, $0.26\%$ and $0.20\%$, respectively \citep{planck2014-a03}.  As noted in Section~\ref{sec:INTRO}, the calibration is absolute in the sense that it is determined from the satellite's orbital motion in the solar system (compared to the speed of light).  It also depends on the absolute temperature of the cosmic microwave background, $T_{o} = 2.7255 \pm 0.0006  K$ \citep{Fixen09}, but the uncertainty in that quantity is at the $0.02\%$ level. 

\subsection{Beam Solid Angles}
These special maps, like all \Planck\ frequency maps, are presented in temperature units.  To convert the measured intensity of a source to flux density, we need to know the size of \Plancks\ effective beam: flux density $S \propto  \Omega \propto ({\rm FWHM})^2$, where $\Omega$ is the entire solid angle of the beam and FWHM is the full width at half maximum, derived from $\Omega$ assuming a Gaussian beam for each receiver. Approximate values for each band are given in Table \ref{tab:PLANCK}.  These were derived from FEBeCoP beams (see \citealt{planck2014-a05} and \citealt{mitra2010}) constructed for these maps; note that the beam shape and solid angle vary slightly from point to point in the sky.  Extensive testing and calculations described in \cite{planck2013-p02d}  give us confidence that we know \Plancks\ beam solid angle in the 30\,GHz channel to a precision of $\sim 0.1\%$.  The situation at 44\,GHz is more complicated.  Two of \Plancks\ three 44\,GHz receiver-horn assemblies are located on one side of its focal plane, and at a substantial distance from its center (see \citealt {planck2011-1.3}; \citealt{planck2014-a05}).   As a consequence, the beams for these two horns are substantially elliptical and are broader than the beam for the third, which is located on the other side of the focal plane.  The FWHM figure in Table \ref{tab:PLANCK} is a weighted average accurate to $\sim 0.2\%$ \citep{planck2014-a05}. The weights used were the same as employed in the LFI mapmaking process \citep{planck2014-a07}.

In Section~ \ref{sec:CompPlanckATCA}, we treat separately the two sets of horns, and the measurements derived from each.  Note that the large separation of the 44\,GHz horns also means that a given source is observed at two separate epochs as the \Planck\ beams scan across the sky; the separation between the two 44\,GHz observations is $\sim ~6$ days.  We return to this issue in Section~\ref{sec:COMP} where we consider the effects of variability in the flux density of these sources.

\subsection{Color Correction and Frequency Interpolation of \Planck\ Measurements \label{sec:CCPlanck}}
Since \Plancks\ calibration is based on the dipole signal induced in the CMB (which has a thermal spectrum with flux density roughly $\propto \nu^2$) but most of the sources in this study have spectral indices $\alpha$ ($S_\nu \propto \nu^{\alpha}$) in a quite different range, 0 to $-1$, the \Planck\ flux densities need to be color-corrected (see \citealt{planck2014-a03}).  This color correction, as well as the (small) interpolation from \Planck\ band centers of 28.4 and 44.1\,GHz to the standard VLA frequencies of 28.45 and 43.34\,GHz, was made individually for each source at each frequency.  The corrections were based on spectral indices derived from the very precise VLA measurements.  To correct the 28\,GHz data, we calculated the spectral index from flux density measurements at 25.836 and 36.435\,GHz, and at 44\,GHz, we used the VLA measurements at 36.435 and 48.425\,GHz.  Typical values of these multiplicative corrections were 0.99--1.005 at 28\,GHz and 0.98--1.00  at 44\,GHz.  For a given source, the color corrections and frequency interpolation can be made with a precision of $\sim0.1\%$ or better.  The color corrections tabulated in \cite{planck2014-a03}, however, have an intrinsic uncertainty of up to $0.4\%$ for the range of spectral indices we find.  These uncertainties are included in our final error budget.

\subsubsection{Extrapolation to VLA and ATCA Frequencies} \label{ExtrapATCA}
As listed in Section~\ref{sec:ATCAObs}, the ATCA frequencies most closely overlapping \Plancks\ LFI bands were 22.45, 43.34, and 48.425\,GHz. As for the VLA, the 43.34\,GHz band was close to the \Planck\ 44.1\,GHz band center.  Thus the color correction and extrapolation required for the \Planck\ measurements were small, as noted above.  As in the case of the VLA results, these corrections were made individually for each source, based on VLA spectral indices when available (and on ATCA 43.34 and 48.425\,GHz measurements otherwise).  Since the ATCA and VLA band centers match exactly, we included all 43.34\,GHz observations made at either instrument when we compare the results to extrapolated \Planck\ measurements.  In taking this step, we assume that the 43.34\,GHz flux density scales at the VLA and ATCA are consistent (confirmed below in Section~\ref{sec:COMPJVLAATCA}).  

We also also combined results from both ground-based instruments at 22.45\,GHz.  Comparing \Planck\ observations with the VLA and ATCA results at 22.45\,GHz, however, requires a much larger extrapolation in frequency, again based on spectral indices determined from VLA or ATCA measurements.  The color correction to some degree cancels the frequency extrapolation, but the overall adjustment required for \Planck\ data ranged from 0.7 to 1.3 (multiplicative).  

\subsection{Resulting \Planck\  Measurements Interpolated to Ground-Based Frequencies}
The interpolated and color-corrected \Planck\ flux densities, adjusted to match the frequencies of the ground-based observations, are given in columns 10, 11, and 13 of Table \ref{tab:fluxes}.  The table also provides the approximate time intervals between \Planck\ and VLA and/or ATCA observations. The tabulated \Planck\ flux densities are averages over the entire three-month period.  In some cases, sources were observed only once, over a period of a day or so; in other cases, especially for sources near the ecliptic poles, sources were observed for several days or more extended periods.  In addition, as noted above, the 44\,GHz measurements for each source occurred at two different epochs.  We consider the effect of these complications further in Section~\ref{sec:COMP}.

\begin{table}
\caption{\Planck\ characteristics. \label{tab:PLANCK}}
\smallskip
\begin{tabular}{lcc}
\tableline\tableline
Band&Band Center Frequency&Beam FWHM\\ 
&(GHz)&(arcmin)\\
\tableline
30    &  28.4     &  32   \\
44    &  44.1    &   27   \\
70    &  70.4    &   13   \\
\tableline
\end{tabular}
\end{table}


\section{Comparison of Flux Densities}\label{sec:COMP}

\subsection{\Planck\ vs. ground-based Flux Density Measurements}
In Figures \ref{fig:JVLA28}, \ref{fig:ATCAJVLA22} and \ref{fig:JVLA44}, we plot the fully corrected \Planck\ flux densities against the corresponding VLA and ATCA measurements as described in Sections~\ref{sec:JVLAObs} and \ref{sec:ATCAObs}.  Not every VLA and ATCA source was detected by \textit{Planck}.  In some cases, this was because the source was too faint (as noted in Section~\ref{sec:INTRO}, sources were selected to cover a range of flux density).  In other cases, at one or another of the \Planck\ frequencies, the source fell just outside the area of the sky \Planck\ scanned in the interval 2013 April 1 to June 30 (see Figure \ref{fig:sky_dist}).  For sources that were observed in both early and late May at the VLA, we treat the two VLA measurements as independent,  and plot both against the \Planck\ results.  

For each frequency, we also plot the best-fit linear relation. Since the \Planck\ flux density errors dominated, and were roughly equal for most sources, we did not weight the data. In addition, we forced these fits to pass through $(0,0)$; these are referred to below as ``constrained fits.'' The consequences of this choice are discussed in Section~\ref{sec:EddBias}.
	
If the \Planck\ and ground-based flux density scales agreed exactly, we would expect an exactly linear relation with unit slope.  Measurement uncertainty in the \Planck\ values (indicated by error bars in the figures), as well as source variability, produces the  scatter seen in the figures. 
	
\subsubsection{Comparison of  \Planck\ with VLA Measurements at 28.45\,GHz}
\begin{figure}[ht]
\begin{center}
\includegraphics[width=0.5\textwidth]{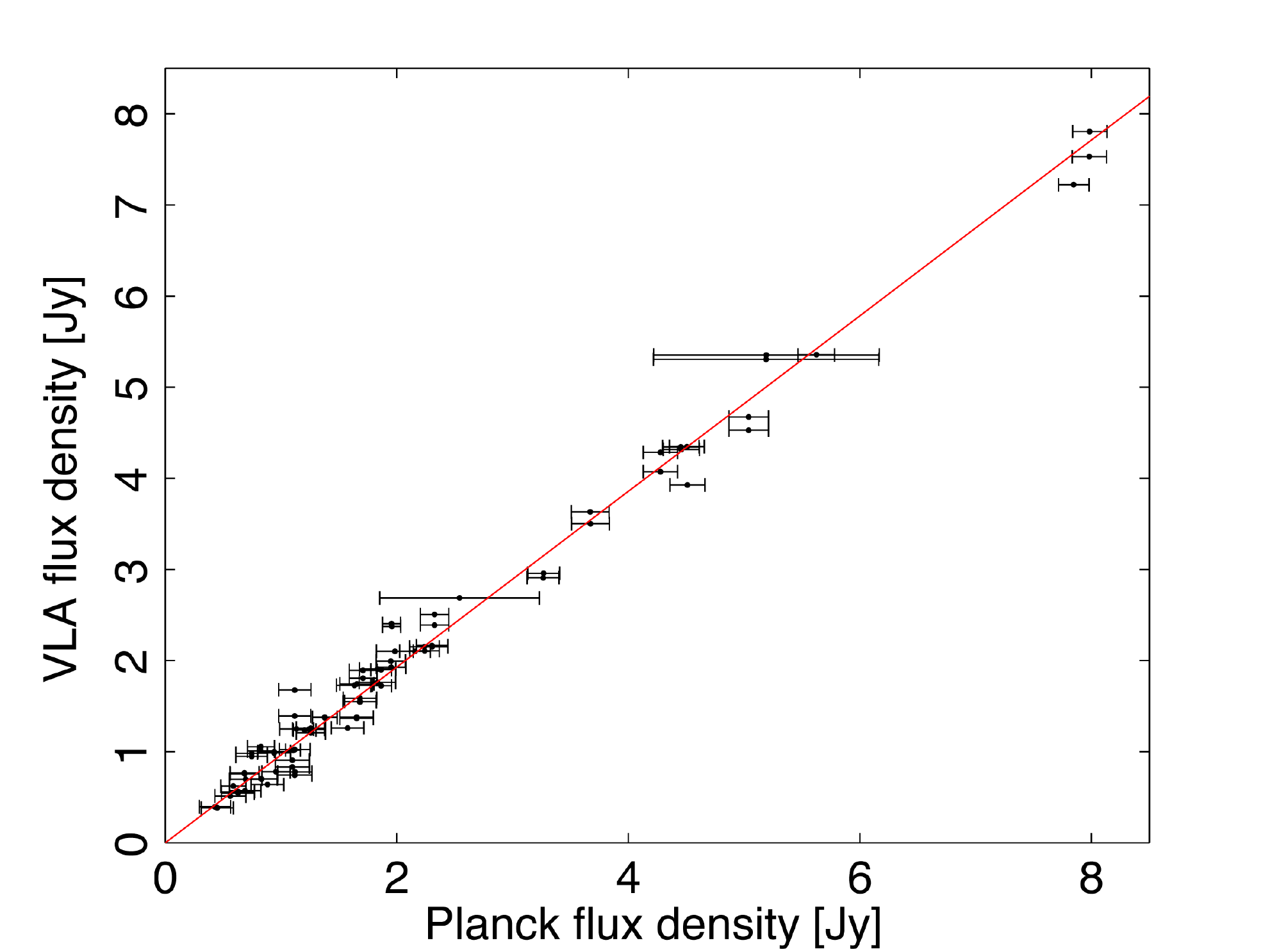}
\caption{Comparison between color-corrected \Planck\ and VLA measurements at 28.45\,GHz; the observed scatter is due mainly to variability of the sources. The slope and $1\sigma$ uncertainty of the fit (solid line) are $ 0.964\pm 0.008$.
\label{fig:JVLA28}}
\end{center}
\end{figure}

Figure \ref{fig:JVLA28} shows the agreement between the corrected \Planck\ and VLA measurements made at 28.45 GHz.  The measured slope and the $1\sigma$ uncertainty of the relation are $ 0.964\pm 0.008$, when the fit is constrained to pass through $(0,0)$. This is changed to $S({\rm VLA}) = 0.948 S(\textit{Planck}) + 0.056$ by allowing an unconstrained fit. We thus find that VLA flux densities run $\sim4\%$ (with a statistical scatter of $\pm 0.8\%$)  below the \Planck\ measurements.  This fit does not take into account any possible systematic errors in the \Planck\ or ground-based measurements.

We next consider the potential systematic uncertainties in the \Planck\ calibration, discussed in detail in \cite{planck2014-a04}. There are three sources of systematic error, which we assume are independent. These include the uncertainty in the color correction mentioned above (taken as $0.4\%$); $0.06\%$ uncertainty in the solid angle of the beam, taken from \cite{planck2014-a05}; and $0.35\%$ calibration uncertainty, taken from \cite{planck2014-a03}.  We combine these systematic errors in quadrature to arrive at an estimate of the systematic error.  VLA flux densities run $3.6\% \pm 0.8\% ({\rm stat}) \pm 0.5\% ({\rm syst})$, or $3.6\% \pm 1.0\%$ lower than those measured by \Planck\, if we combine the two types of error in quadrature. This result is cited in \cite{planck2014-a35}.  Estimates of the systematic uncertainties in the VLA and ATCA flux densities are given in sections 2 and 3.2.4.

\subsubsection{Comparison of \Planck\ with VLA and ATCA Measurements at 22.45\,GHz} \label{sec:CompPlanckATCA}
As noted in Section~\ref{ExtrapATCA}, comparing \Planck\ measurements made at 28.45\,GHz to VLA and ATCA values at 22.45\,GHz required a much larger extrapolation in frequency. This extrapolation (like the generally smaller color correction) made use of the spectral index for each source.  We employed several means of calculating spectral indices for each source. First, for those sources observed by the VLA, the spectral index could be found directly from 22.45 and 28.45\,GHz observations.  For sources observed only at the ATCA, we calculated both a spectral index from 17 to 22\,GHz and one from 22 to 43\,GHz.  For all but four of the sources involved, all the spectral indices agreed within errors, and we used an average for the extrapolation and color correction. 

\begin{figure}[ht]
\begin{center}
\includegraphics[width=0.5\textwidth]{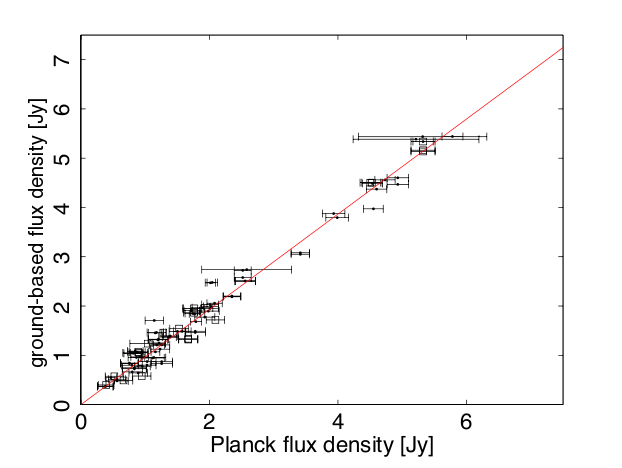}
\caption{Color-corrected and extrapolated \Planck\ flux densities compared to VLA measurements (\textit{dots}) and ATCA measurements (\textit{open squares}) of the same sources at 22.45\,GHz; The slope and $1\sigma$ uncertainty of the fit (solid line) are $ 0.967\pm 0.007$.
\label{fig:ATCAJVLA22}}
\end{center}
\end{figure}

For the other four cases, we compared results using the largest value for the derived spectral index for a given source with the results with the smallest value.  This resulted in a $\sim 1\sigma$ shift in the overall slope.  The results we adopted are based on taking that value for the spectral index of these 4 sources which produced the lowest scatter in the fit. As shown in Figure  \ref{fig:ATCAJVLA22} , we find a slope of $0.967 \pm 0.007 ({\rm stat})$.  In the case of the 22\,GHz measurements, ground-based flux densities again run low, but agree within the assumed ATCA and VLA uncertainties with the \Planck\ values.

\subsubsection{Comparison of Measurements at 43\,GHz}
\begin{figure}[ht]
\begin{center}
\includegraphics[width=0.488\textwidth]{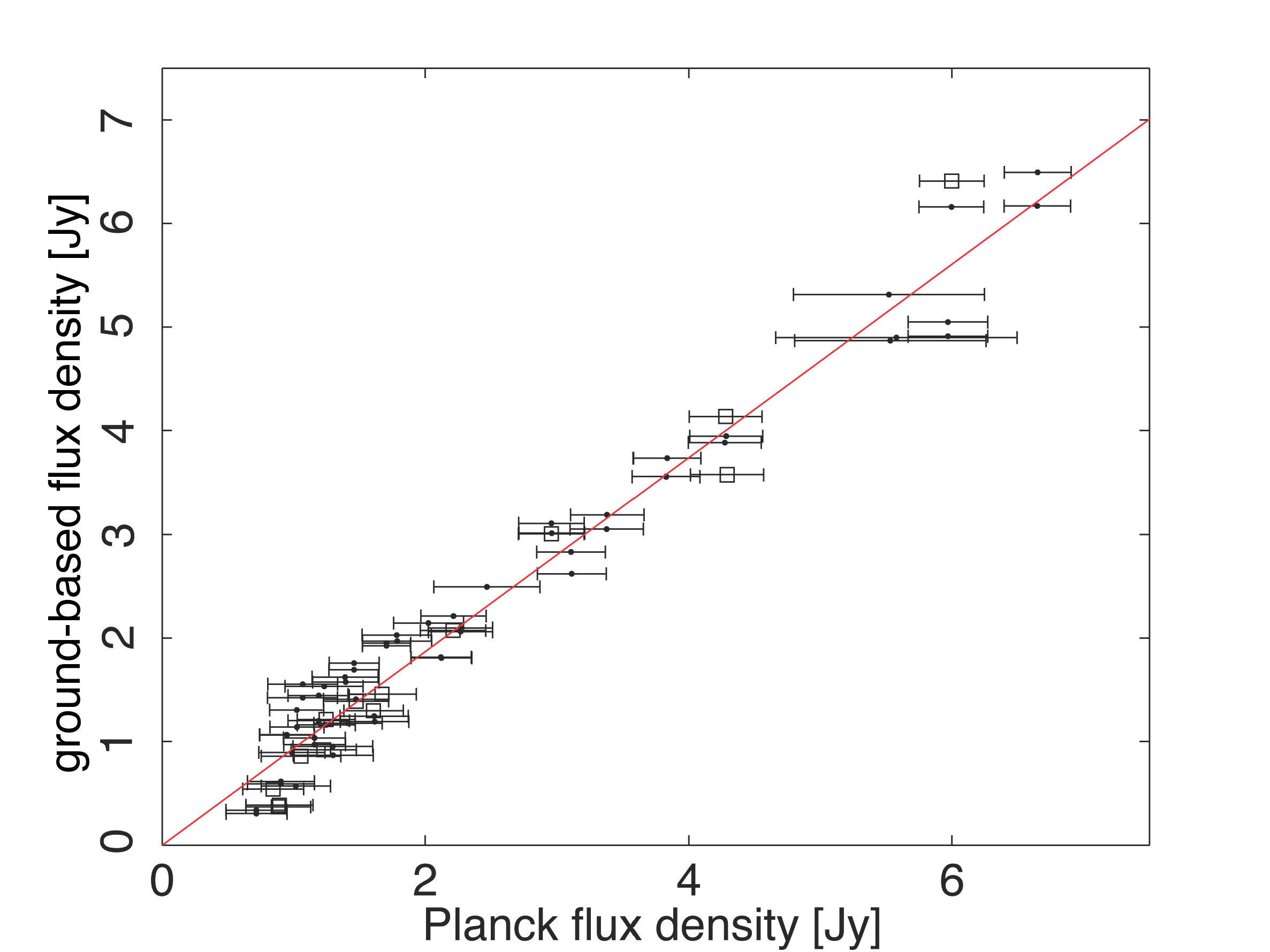}
\caption{Color-corrected (and extrapolated) \Planck\ flux densities compared to VLA measurements (\textit{dots}) and ATCA measurements (\textit{open squares}) of the same sources at 43.34\,GHz. The slope and $1\sigma$ uncertainty of the constrained fit (solid line) are $ 0.9384\pm 0.013$. 
\label{fig:JVLA44}}
\end{center}
\end{figure}

 Figure \ref{fig:JVLA44} demonstrates the agreement between the flux density scales used by \Planck\ and by the two ground-based instruments at 43.34\,GHz.  The constrained linear fit to all the data shows that the ground-based measurements are on average $6.2 \% \pm 1.3 \%$ lower than \Planck's. Allowing an unconstrained fit gives $S({\rm ground~based)} = 0.933S(\textit{Planck}) + 0.018$.  While the extrapolation from the \Planck\ band center of 44.1\,GHz to the VLA and ACTA frequency of 43.34\,GHz is slightly larger than at 28.45\,GHz, it is partially offset by the spectral-index dependent color correction.  At 43.34\,GHz, the overall uncertainty is dominated by the $1.3\%$ statistical error in the slope of the fit, induced by \Planck\ measurement errors and variability.   

We estimate the systematic error as the quadrature sum of the uncertainty in the color correction $(0.4\%)$, the beam uncertainty $<0.2\%$ \citep{planck2014-a05}, and the overall calibration $<0.26\%$ \citep{planck2014-a03}. We thus end with an observed difference in the flux density scales of $6.2\% \pm 1.3\% ({\rm stat}) \pm 0.5\% ({\rm syst})$; the ground-based measurements are fainter than \Planck\ values by $6.2\% \pm 1.4 \%$.  

This discrepancy is much larger than either the color correction or the uncertainty in \Plancks\ beam solid angle, and also exceeds the quoted $3\%$ accuracy of the flux density scale introduced by \cite{Perley13}.  The discrepancy between \Planck\ and ground-based flux densities is not significantly changed if we consider only the VLA measurements: if we omit the smaller number of ATCA measurements, the slope changes from 0.9384 to 0.9343.  

\subsubsection{Treating the 44\,GHz Horns Separately}
To explore this discrepancy further, we also considered separately the \Planck\ measurements made with the single 44\,GHz horn on one side of the focal plane, and measurements made by the two other horns on the other side of the focal plane. The separate measurements were noisier.  We found a $1.7\sigma$ difference: the flux densities recorded by the single horn for these sources were on average $4.6\% \pm 2.7\%$ higher.  Even if we exclude all measurements made by this horn, however, we still find that the remaining \Planck\ measurements at 44\,GHz run high compared to ground-based ones.

\subsection{The Effect of Resolved Sources and Possible Confusion}\label{sec:EffResol}
The VLA beams have far smaller solid angles than \textit{Planck's}.  As a consequence, four sources were heavily resolved by the VLA but not by \textit{Planck}: J0813+4812 (3C\,196), J1229+0203 (3C\,273), J1411+5212 (3C\,295) and J2107+4213 (the planetary nebula NGC\,7027).  Although in comparing measurements from the two instruments we always used the total VLA flux density, which nominally corrects for resolution, we examined the effect that dropping these four sources had on the slopes of the fit. If some flux were missed at the VLA due to resolution, we would expect the slopes to increase slightly when dropping these sources.  At 43.34\,GHz, dropping these sources had only a small effect on the slope: it did increase slightly to $0.941 \pm 0.013$. At 28.45\,GHz, the effect of dropping resolved sources was equally small, but in the opposite direction: the slope decreased to $0.958 \pm 0.008$. The same was true at 22.45\,GHz: the slope changed to $0.963 \pm 0.007$.

We also investigated the possibility that \Plancks\ larger beam could incorporate radio sources other than the target source (loosely, "confusion").  To first order, the Mexican Hat Wavelet 2 algorithm used to derive \Plancks\ flux densities corrects for a random distribution of weak sources. In addition, we computed the probability of finding a weak source within \Plancks\ 32 arcminute beam at 28\,GHz, employing 30 GHz source counts taken from \cite{planck2011-6.1}.  That probability falls below unity for sources with flux density $>10\,mJy$, which in turn is 2.5\% or less of the flux density of  even the weakest target sources we consider. At 44\,GHz the counts (ibid) are lower and the beam solid angle is somewhat smaller, so the probability of finding a source $>10\,mJy$ falls to 20\%.

A referee pointed out that the distribution of radio sources near our bright target sources might not be random.  Radio sources, unlike dusty galaxies, are only weakly correlated -- see, for example, \cite{Cress96} -- so our assumption of a random distribution is not unreasonable.  Nevertheless, we performed a check by searching the NVSS catalog \citep{Condon98} for other sources within 10 arcmin of our target sources.  This catalog is constructed at 1.4\,GHz a much lower frequency than we employed, but covers much of the sky.  Not surprisingly, given the well-established source counts at 1.4\,GHz, we found a few weak sources around many of the target sources (the number ranged from 0 to 8).  We also checked 20 random positions and found comparable numbers of weak sources in the same search area.  Thus we found no indication of a significant increase in the number of weak radio sources near our bright target sources.  Furthermore, none of the weak sources except one near J1037-2934 had a measured 1.4\,GHz flux density $>6\%$ of the flux density of the target source in that search area.  While the flux densities of most of these sources are unknown at the higher frequencies of \Planck\, given typical synchrotron spectral indices, we expect them to be 4-30 times lower at 22 to 44\,GHz.  We end by noting that in the particular case of J1037-2934, the \Planck\ flux densities were observed to be slightly below, not above, the VLA values.

\subsection{The Effect of Source Variability}\label{sec:source_var}
Although we aimed to make the \Planck\ and ground-based observations as close in time as possible, all the VLA observations were made at just two epochs in May, and the ATCA observations ended in late April.  As columns 14 and 15 of Table \ref{tab:fluxes} show, the intervals between ground and space-based measurements ranged from less than a day to 45 days as a consequence.  Since many of our sources are AGN (including many blazars) we expect them to vary.  Variability will introduce scatter into our plots, but should not in principle bias them.  

\subsubsection{Possible effects of bias} \label{sec:EddBias}
One could argue, however, that some form of selection bias would make it less likely for \Planck\ to detect marginal sources when they happen to be in their low luminosity states, thus artificially boosting the average \Planck\ flux densities at the faint end.  Indeed, there is evidence in Figure~\ref{fig:JVLA44} that the 44\,GHz \Planck\ measurements at the faint end may be subject to an effect that biases the \Planck\ flux density measurements high for the faintest sources (see the detailed discussion in \citealt{Crawford10}).  We reduce the impact of this effect by forcing the fits to pass through $(0,0)$.  We also made a trial of dropping all 44\,GHz sources with \Planck\ flux densities $< 1.4$\,Jy.  The result was to shift the slope of the constrained fit slightly to $0.943 \pm 0.013$.   The unconstrained fit to the strong ($S > 1.4$\,Jy) 44\,GHz sources has a much flatter slope of 0.87.   Bias in the flux densities of weak sources is evidently not responsible for the discrepancy between Planck and ground-based measurements at 43\,GHz. The closer spacing in time of the ATCA observations made such a test less useful for those data. 

\subsubsection{Dropping known variable or resolved sources}
As one way of assessing the effect of source variability on our comparisons, we looked first at the sources (about half the sample) observed twice at the VLA, once in early May and once at the end of the month.  At 43\,GHz, five sources changed observed flux density by more than $\sim6\%$ over that interval:  J0813+4812 (resolved), J0958+6533, J1037$-$2914, J2107+4213  (NGC\,7027, resolved), and J2146$-$1525.  In addition, J0854+2006 and J1849+6705 varied at 28\,GHz.  Note that two of these apparently "variable" sources J0813+4812 and J2107+4213  are among the four resolved sources discussed in Section~\ref{sec:EffResol} above. Although we always employed total flux densities, the apparent change in VLA flux density of the two heavily resolved sources is certainly in large part due to the different beam solid angles resulting from the different configurations used in early and late May.  At 43.34\,GHz, as these variable sources were dropped one by one, the slope settled to  $0.941 \pm 0.013$.  As expected, variability introduces scatter but no significant bias in the fits. J2253+1608 = 3C\,454.3 is a special case.  While the observed change in flux density between the two epochs of the VLA observations happened to be small at all frequencies, the source clearly did vary during the (longer) \Planck\ mission  \citep{planck2011-6.3a}.  VLA and \Planck\ measurements at 43\,GHz differ by 15--18\%.  If we also drop this source on grounds of assumed variability, the derived slope at 43\,GHz moves to $0.958 \pm 0.013$, a $\sim1.5\sigma$ change. 

Thus dropping variable and/or resolved sources reduces, but does not eliminate, the discrepancy between \Planck\ and ground-based flux densities at 43\,GHz. 

\begin{figure}[t]
\begin{center}
\includegraphics[width=0.5\textwidth]{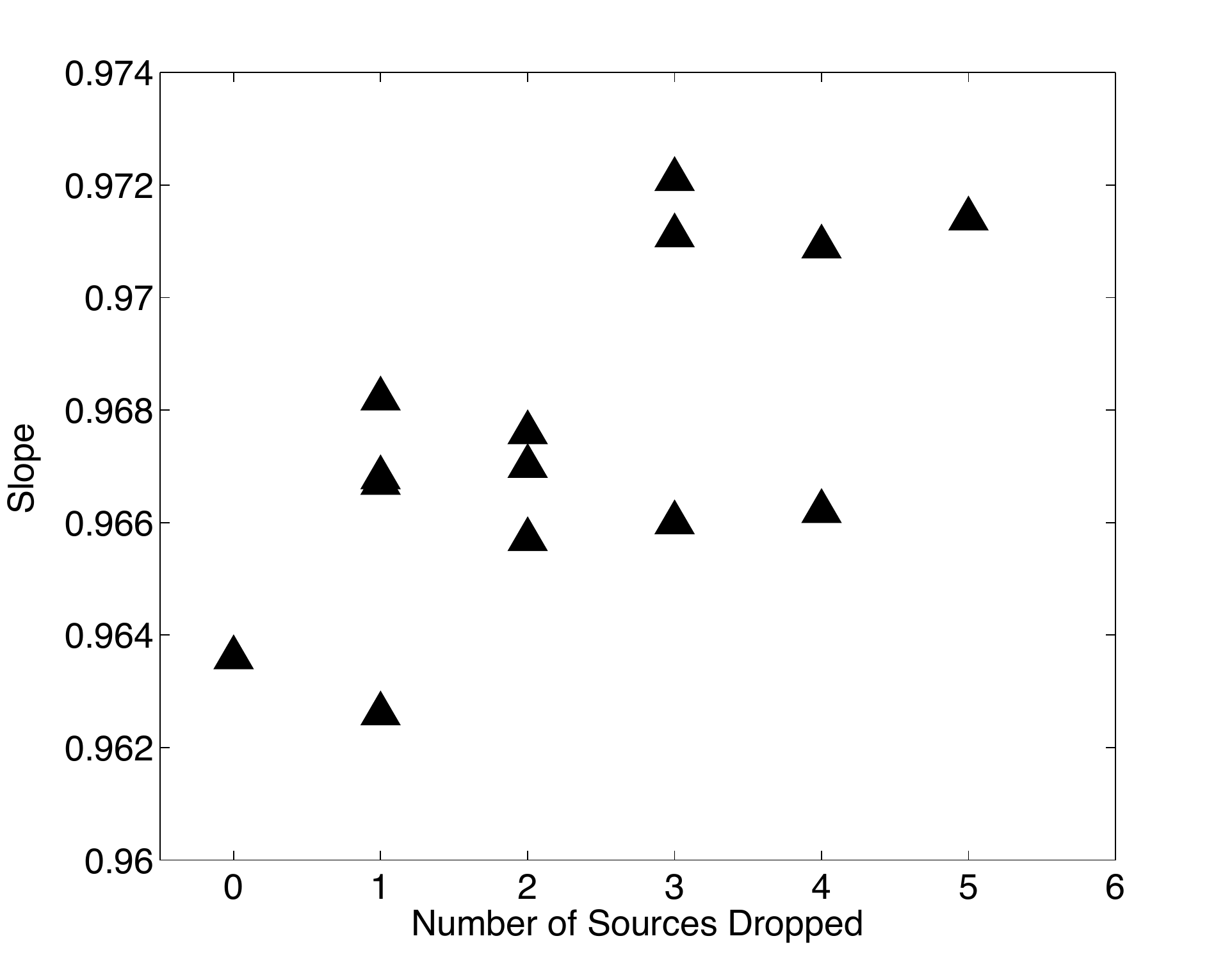}
\caption{Change in the slope of the constrained fits of the 28.45\,GHz data as more and more variable sources are dropped (in various orders); note that the resulting change is smaller than or comparable to the statistical uncertainty in the slope of $\pm 0.008$, and that the slope settles to $\sim 0.97$.  
\label{fig:SLOPE}}
\end{center}
\end{figure}

In contrast, a more systematic change was found at 28.45\,GHz as variable sources (five in this case) were dropped one by one. In Figure~\ref{fig:SLOPE}, we show the result of dropping more and more sources (in various orders): the slope averaged to $0.97 \pm 0.0075$. Again, J2253+1608 had a particularly large effect; also dropping just that source produced a $\sim1\sigma$ change in the slope to $0.977 \pm 0.0075$.  On the other hand, dropping variable and/or resolved sources had little effect on the 22\,GHz data: the slope remained in the range 0.963 to 0.972.

\subsubsection{Restricting the allowed time interval between \Planck\ and ground-based observations}
We also tried restricting the fits to sources observed by \Planck\ within one week of one of the VLA or ATCA runs.  That left only about half of the data. As expected, this restriction reduced the scatter in the plots of ground-based vs.\ \Planck\ flux density.  At 28.45\,GHz, the slope changed by approximately $1\sigma$ to $0.980 \pm 0.009$.  At 43.34\,GHz, the same restriction led to a change in slope from $ 0.9384 \pm 0.013$ to $0.940 \pm 0.012$ (note that the uncertainty barely changed even though we retained only about half the data, because the scatter in the data induced by variability was reduced).

Given these results, we conclude that variability does not alter the overall conclusion that the ground-based flux density scales run below those established by \textit{Planck}.  The ground-based flux densities run roughly 3\% below the \Planck\ flux densities at 22 and 28\,GHz, and 5--6\% below them at 43\,GHz.  

\subsection{The Effect of Color Corrections}
As noted in Section~\ref{sec:CCPlanck}, the \Planck\ data needed to be color corrected as well as interpolated to match the VLA and ATCA central frequencies. We performed two tests on the effect of the small color corrections made to the \Planck\ data.  First, we compared the uncorrected \Planck\ observations in the 30\,GHz band to the VLA results.  Omitting the color corrections changed the observed slope from 0.964 to 0.968.  We also separated the \Planck\ data into two parts, one having a very small $(1.000 \pm 0.005)$ color correction, the other with a larger and more spectral-index-dependent color correction.  The difference in the resulting fits to VLA data was at the $0.1\sigma$ level.

\subsection{Comparison of VLA and ATCA Flux Density Scales at 43.34\,GHz} \label{sec:COMPJVLAATCA}
 We turn next to a direct comparison of flux density scales employed by the two major interferometric arrays, the VLA in the north and the ATCA in the south. We note that questions have been raised   (see \citealt{Sajina11}) about the agreement of calibration scales at 22\,GHz, before the recent re-estimates of flux density scales described in Sections~2 and 3 above. These direct comparisons between the new Perley-Butler and the new Stevens scales are treated in \cite{Stevens15} and cover a much wider range of frequencies than those considered here, as well as a wider range of flux density. Since we have combined 43.34\,GHz observations by the two instruments when making the comparison to \Planck,\ however, we need to be certain that the flux density scales used at the two interferometers agree at that frequency.  We checked this by directly comparing VLA and ATCA measurements for 14 sources observed in common at 43.34\,GHz: on average, VLA measurements were $0.990 \pm 0.011$ those of ATCA measurements of the same source -- at the highest frequencies, the new Perley-Butler and Stevens flux density scales agree well. 

\subsection{Best estimates of VLA--\Planck\ flux density comparison}
We summarize this section by giving our best estimates of the small difference between flux densities observed by \Planck\ and the ground-based instruments.  We take account of source variability as discussed in Section~\ref{sec:source_var}.  At 28.45\,GHz, we find that VLA flux densities are lower than  \textit{Planck}'s by 2--3\% with a combined systematic and statistical error of $1\%$.  At 22.45\,GHz, ATCA and VLA flux densities are 3--3.5\% lower than \textit{Planck}'s, again with a combined error of $\sim1\%$.
   At 43.34\,GHz, on the other hand, the discrepancy is larger: \Planck\ flux densities are higher by 5--6\% than those from either ground-based instrument, with a combined systematic and statistical error of 1.4\%. At both 28.45 and 43.34\,GHz, these results are consistent with, but more precise than, the values based on a less complete analysis of these data in \cite{planck2014-a35}.

\section{Polarization} \label{sec:POL}
We attempted to compare \Planck\ measurements of polarized flux density with those  obtained at the VLA and ACTA.  Very few of the sources observed in this program had enough polarized flux to be robustly detected by \textit{Planck}. One of these was 3C\,273.  The \Planck\ 30\,GHz polarized flux density (color-corrected and extrapolated to 28.45\,GHz for comparison with the VLA) was $835 \pm 70$ mJy, while the VLA observed $813 \pm 70$ mJy.  At 43.34\,GHz the \Planck\ polarized flux density was $567 \pm 97$ mJy, as compared with $623 \pm 70$ mJy seen by the VLA.  At both frequencies, the measured polarization angles agreed to $\pm 2\deg$. Further refinements of the \Planck\ polarization measurements are planned, including more scrutiny of the possible leakage of total power into polarization.  At this point, we simply report that there are no evident discrepancies between the \Planck\ and ground-based polarization.

\section{Summary and Discussion} \label{sec:conclusions}
In conclusion, we see that the comparison of \Planck\ data with VLA data at 28.45\,GHz yields acceptable agreement: the VLA data run on average 2--3\% fainter than \Planck's, which is within the margin of error of the VLA flux density scale.  Similarly, the \Planck\ 28.45\,GHz data that have been extrapolated and compared to the VLA and ATCA data at 22.45\,GHz are about $3-3.5\%$ higher than the ground-based measurements, which is just within the margin of statistical and systematic error. The more problematic comparison is that between \Planck\ 44.1\,GHz data (extrapolated to 43.34\,GHz) and the ground-based instruments at 43.34\,GHz.  \Planck\ data were consistently 5--6\% higher than the ground-based measurements. This remained largely unchanged when the \textit{Planck} data were compared with the VLA and ATCA individually, when the two sets of 44\,GHz \Planck\ horns were compared separately to the ground-based instruments, when four resolved sources from the VLA data were dropped, and when variable sources were dropped (individually and in all combinations). We suggest therefore that the difference in flux measurements in the 43\,GHz range results from a difference in calibration. This discrepancy could affect precision flux density comparisons between instruments, and would also affect source spectra that included 43\,GHz data.

We end by asking whether the difference in flux density scales we have demonstrated could be due to a calibration mismatch between \Planck\ and \textit{WMAP}, since the VLA flux density scale is based on the \textit{WMAP} calibration.  \Planck--\textit{WMAP} calibration has been examined in detail in \cite{planck2014-a03} and \cite{planck2014-a06} for the \Planck\ bands considered here.  Following the initial release of \Planck\ data in 2013, there was a small upward adjustment of the \Planck\ calibration (which brings it closer to the \textit{WMAP} calibration).  The shifts were $+0.45 \%$ and $+ 0.64 \%$  at 28 and 44\,GHz, respectively.  Even with these slight shifts, the \textit{WMAP} calibration remains $\sim1\%$ higher than \Plancks, so the difference is unlikely to explain why the VLA flux densities appear to be a bit low.  It is important to note, however, that the comparison of \Planck\ and \textit{WMAP} calibration is based on observations of the CMB (both the dipole and CMB fluctuations at degree scales).  These have a different spectrum than the radio sources considered here, and have larger angular scales.  A better comparison is between \Planck\ and \textit{WMAP} observations of planets (essentially point sources for both instruments).  Measurements of Jupiter are compared in \cite{planck2014-a06}.  These indicate that \Plancks\ measurements of the brightness temperature of Jupiter agree with \textit{WMAP}'s to $0.2 \pm 1.0\%$ at 28\,GHz, and $\sim 0.0 \pm 1\%$ at 44\,GHz.  Thus we currently have no convincing explanation for the observed discrepancy. We can speculate on possible contributing factors: errors in the model for secular changes in the emission of Mars, or in corrections for atmospheric absorption, or in the beam solid angles of one of the satellite experiments.  Further refinements of the \Planck\ data and analysis, expected in the next year, may allow us to check the last of these.

\begin{acknowledgements}
We are deeply indebted to Kris Gorski, Sanjit Mitra, and Luca Pagano of the FEBeCoP team who helped in the construction of the beams used to derive flux densities from the \Planck\ maps.  The dates that \Planck\ observed each source were supplied by Jonathan Leon Tavares, then at Aalto University, Finland. MLC acknowledges the Spanish MINECO Projects AYA2012-39475-C02-01 and Consolider-Ingenio 2010 CSD2010-00064. The \Planck\ Collaboration acknowledges the support of: ESA; CNES and CNRS/INSU-IN2P3-INP (France); ASI, CNR, and INAF (Italy); NASA and DoE (USA); STFC and UKSA (UK); CSIC, MINECO, JA, and RES (Spain); Tekes, AoF, and CSC (Finland); DLR and MPG(Germany); CSA (Canada); DTU Space (Denmark); SER/SSO (Switzerland); RCN (Norway); SFI (Ireland); FCT/MCTES (Portugal); ERC and PRACE (EU).  A description of the \Planck\ Collaboration and a list of its members, indicating which technical or scientific activities they have been involved in, can be found at http://www.cosmos.esa.int/web/planck/planck-collaboration.
\end{acknowledgements}

{\it Facilities:} \facility{ATCA}, \facility{Planck}, \facility{VLA}

\bibliographystyle{apj}
\bibliography{Planck_bib,planck_vla}

\clearpage
\begin{deluxetable}{lccccccccccccccc}
\rotate
\tabletypesize{\scriptsize}
\tablecaption{Flux Density Measurements\label{tab:fluxes}}
\tablehead{\colhead{} & \colhead{1} & \colhead{2} & \colhead{3} & \colhead{4} & \colhead{5} & \colhead{6} & \colhead{7} & \colhead{8} & \colhead{9} & \colhead{10} & \colhead{11} & \colhead{12} & \colhead{13} & \colhead{14}  & \colhead{15}  \\ 
\colhead{} & \colhead{VLA} & \colhead{VLA} & \colhead{VLA} & \colhead{VLA} & \colhead{VLA} & \colhead{VLA} & \colhead{ATCA} & \colhead{ATCA} & \colhead{Planck} & \colhead{Planck} & \colhead{Planck} & \colhead{Planck} & \colhead{Planck} & \colhead{Time Gap} & \colhead{Time Gap}  \\ 
\colhead{Source} & \colhead{22.450} & \colhead{25.836} & \colhead{28.450} & \colhead{36.435} & \colhead{43.340} & \colhead{48.425} & \colhead{22.45} & \colhead{43.34} & \colhead{Raw 28} & \colhead{Corr. 28} & \colhead{Corr. 22} & \colhead{Raw 44} & \colhead{Corr 43} & \colhead{30\,GHz} & \colhead{44\,GHz}  \\ }
\startdata
J0725$-$0054  & 5.4412 &  5.3963 &  5.3553 &  5.2238 &  4.9889 &  4.9017 &  \ldots &  \ldots &  5.6118 &  5.6138 &  5.7801 &  \ldots &  \ldots &  28 &  \dots \\ 
J0745$-$0044  & 1.3696 &  1.2990 &  1.2463 &  1.1203 &  1.0332 &  0.9863 &  \ldots &  \ldots &  1.241 &  1.2387 &  1.3651 &  \ldots &  \ldots &  22 &  21 \\ 
  & \dots  & \ldots &  \ldots &  \ldots &  \ldots &  \ldots &  \ldots &  1.0190 &  1.241 &  \ldots &  1.3656 &  \ldots &  \ldots &  19 &  18 \\ 
B0754+100  & \ldots & \ldots &  \ldots &  \ldots &  \ldots &  \ldots &  1.0680 &  0.9200 &  0.8897 &  \ldots &  0.8913 &  1.2336 &  1.2282 &  6 &  11 \\ 
J0813+4812  & 0.7266 &  0.6345 &  0.5461 &  0.3995 &  0.3064 &  0.2546 &  \ldots &  \ldots &  0.6356 &  0.6289 &  0.8326 &  0.7133 &  0.7180 &  44 &  20 \\ 
  & 0.7452 &  0.6243 &  0.5592 &  0.4043 &  0.3365 &  0.2801 &  \ldots &  \ldots &  0.6356 &  0.6295 &  0.8326 &  0.7133 &  0.7166 &  18 &  \ldots \\ 
J0826$-$2230  & 1.3345 &  1.3029 &  1.2601 &  1.2051 &  1.1611 &  1.1342 &  \ldots &  \ldots &  1.5763 &  1.5758 &  1.6708 &  1.2537 &  1.2482 &  3 &  14 \\ 
  &  \ldots & \ldots &  \ldots &  \ldots &  \ldots &  \ldots &  1.3190 &  1.2130 &  1.5763 &  \ldots &  1.6708 &  1.2537 &  1.2470 &  \ldots &  1 \\ 
B0826$-$373  & \ldots &  \ldots &  \ldots &  \ldots &  \ldots &  \ldots &  1.2360 &  0.8600 &  0.9241 &  \ldots &  1.0399 &  1.0575 &  1.0560 &  \ldots &  \ldots \\ 
B0829+046  & \ldots & \ldots &  \ldots &  \ldots &  \ldots &  \ldots &  0.5740 &  0.5420 &  0.5147 &  \ldots &  0.5203 &  0.8488 &  0.8442 &  3 &  1 \\ 
B0834$-$201  & \ldots & \ldots &  \ldots &  \ldots &  \ldots &  \ldots &  1.7180 &  1.2980 &  1.8841 &  \ldots &  2.0921 &  1.6113 &  1.6075 &  \ldots &  1 \\ 
J0854+2006  & 3.9721 &  4.0049 &  3.9266 &  3.8965 &  3.8849 &  3.9594 &  \ldots &  \ldots &  4.5059 &  4.5075 &  4.5509 &  4.3026 &  4.2730 &  27 &  27 \\ 
  & 4.5606 &  4.4371 &  4.3473 &  4.1311 &  3.9458 &  3.8957 &  \ldots &  \ldots &  4.5059 &  4.5039 &  4.7311 &  4.3026 &  4.2838 &  1 &  1 \\ 
  & \ldots &\ldots &  \ldots &  \ldots &  \ldots &  \ldots &  4.5030 &  4.1350 &  4.5059 &  \ldots &  4.5008 &  4.3026 &  4.2795 &  8 &  2 \\ 
  & \ldots &\ldots &  \ldots &  \ldots &  \ldots &  \ldots &  4.5030 &  3.5740 &  4.5059 &  \ldots &  4.5277 &  4.3026 &  4.2902 &  8 &  14 \\ 
J0900$-$2808  & 0.5523 &  0.5263 &  0.5080 &  0.4552 &  0.3936 &  0.3698 &  \ldots &  \ldots &  \ldots &  \ldots &  \ldots &  \ldots &  \ldots &  \ldots &  11 \\ 
  & \dots &\ldots &  \ldots &  \ldots &  \ldots &  \ldots &  \ldots &  0.4080 &  \ldots &  \ldots &  \ldots &  \ldots &  \ldots &  \ldots &  \ldots \\ 
J0909+0121  & 2.0568 &  2.1213 &  2.1504 &  2.1301 &  2.0977 &  2.0489 &  \ldots &  \ldots &  2.2373 &  2.2394 &  2.0806 &  2.2785 &  2.2674 &  17 &  1 \\ 
  & 1.9910 &  2.0569 &  2.1077 &  2.1124 &  2.0615 &  2.0021 &  \ldots &  \ldots &  2.2373 &  2.2399 &  2.0247 &  2.2785 &  2.2685 &  1 &  \ldots \\ 
J0920+4441  & 1.3263 &  1.3010 &  1.2477 &  1.1419 &  1.0669 &  1.0525 &  \ldots &  \ldots &  1.1352 &  1.1339 &  1.2089 &  0.9517 &  0.9485 &  25 &  1 \\ 
  & 1.3150 &  1.2583 &  1.2473 &  1.1183 &  1.0637 &  0.9863 &  \ldots &  \ldots &  1.1352 &  1.1339 &  1.1976 &  0.9517 &  0.9499 &  1 & \ldots   \\ 
J0927+3902  & 8.1303 &  7.7399 &  7.5306 &  6.8295 &  6.1684 &  5.9942 &  \ldots &  \ldots &  7.9907 &  7.9816 &  8.6299 &  6.6606 &  6.6481 &  22 &  1 \\ 
  & 8.3966 &  7.9208 &  7.8075 &  7.0650 &  6.4944 &  6.0536 &  \ldots &  \ldots &  7.9907 &  7.9824 &  8.6299 &  6.6606 &  6.6514 &  1 &  \ldots  \\ 
J0927$-$2034  & 2.0287 &  2.0854 &  1.9939 &  1.8813 &  1.8093 &  1.6940 &  \ldots &  \ldots &  1.9514 &  1.9500 &  2.0099 &  2.1300 &  2.1239 &  2 &  3 \\ 
  & 1.9501 &  1.9466 &  1.9256 &  1.8807 &  1.8155 &  1.7674 &  \ldots &  \ldots &  1.9514 &  1.9507 &  2.0099 &  2.1300 &  2.1207 &  10 &  \ldots  \\ 
J0948+4039  & 0.8293 &  0.7814 &  0.7433 &  0.6495 &  0.5917 &  0.5473 &  \ldots &  \ldots &  1.1227 &  1.1202 &  1.2574 &  0.9040 &  0.9037 &  17 &  1 \\ 
  & 0.8708&  0.8103 &  0.7767 &  0.6831 &  0.6150 &  0.5750 &  \ldots &  \ldots &  1.1227 &  1.1203 &  1.2574 &  0.9040 &  0.9037 &  2 &  \ldots  \\ 
J0956+2515  & 1.4862 &  1.5538 &  1.5868 &  1.6306 &  1.6225 &  1.7411 &  \ldots &  \ldots &  1.6787 &  1.6808 &  1.5024 &  1.4048 &  1.3916 &  6 &  9 \\ 
  & 1.4884 &  1.5178 &  1.5456 &  1.5740 &  1.5744 &  1.5790 &  \ldots &  \ldots &  1.6787 &  1.6808 &  1.5779 &  1.4048 &  1.3959 &  9 &  \ldots  \\ 
J0958+6533  & 1.2091 &  1.2440 &  1.2564 &  1.2638 &  1.3060 &  1.2690 &  \ldots &  \ldots &  1.2575 &  1.2586 &  1.1820 &  1.0302 &  1.0236 &  27 &  1 \\ 
  & 1.2118 &  1.1983 &  1.2069 &  1.1726 &  1.1414 &  1.1085 &  \ldots &  \ldots &  1.2575 &  1.2584 &  1.2575 &  1.0302 &  1.0257 &  2 &  \ldots \\ 
J1037$-$2934  & 1.4619 &  1.4299 &  1.3915 &  1.2539 &  1.2015 &  1.1026 &  \ldots &  \ldots &  1.1202 &  1.1187 &  1.1762 &  1.1931 &  1.1909 &  \ldots &  \ldots \\ 
  & 1.7056 &  1.6865 &  1.6775 &  1.5661 &  1.4445 &  1.3955 &  \ldots &  \ldots &  1.1202 &  1.1202 &  1.1426 &  1.1931 &  1.1903 &  \ldots &  \ldots \\ 
J1044+8054  & 1.0657&  1.0212 &  1.0067 &  0.9262 &  0.8760 &  0.8420 &  \ldots &  \ldots &  0.8271 &  0.8265 &  0.8767 &  \ldots &  \ldots &  \ldots &  \ldots \\ 
  & 1.0851&  1.0557 &  1.0543 &  0.9901 &  0.9279 &  0.8876 &  \ldots &  \ldots &  0.8271 &  0.8270 &  0.8477 &  \ldots &  \ldots &  \ldots &  \ldots \\ 
J1048+7143  & 1.8504 &  1.8296 &  1.8064 &  1.7218 &  1.6938 &  1.6419 &  \ldots &  \ldots &  1.7089 &  1.7087 &  1.7430 &  1.4655 &  1.4584 &  24 &  13 \\ 
  & 1.9063 &  1.8779 &  1.8911 &  1.8215 &  1.7574 &  1.6954 &  \ldots &  \ldots &  1.7089 &  1.7098 &  1.7259 &  1.4655 &  1.4598 &  1 &  \ldots \\ 
J1048$-$1909  & 1.8979 &  1.8446 &  1.7597 &  1.5760 &  1.4246 &  1.4053 &  \ldots &  \ldots &  1.8368 &  1.8332 &  1.9837 &  1.0704 &  1.0679 &  \ldots &  \ldots \\ 
  & 1.9657&  1.9098 &  1.9078 &  1.7110 &  1.5557 &  1.4863 &  \ldots &  \ldots &  1.8368 &  1.8356 &  1.8919 &  1.0704 &  1.0689 &  \ldots &  \ldots \\ 
J1104+3812  & 0.6433 &  0.6400 &  0.6384 &  0.6364 &  0.6349 &  0.6275 &  \ldots &  \ldots &  0.8833 &  0.8841 &  0.8921 &  \ldots &  \ldots &  \ldots &  \ldots \\ 
J1130+3815  & 1.1229 &  1.0702 &  1.0232 &  0.9357 &  0.8961 &  0.8480 &  \ldots &  \ldots &  1.1205 &  1.1188 &  1.2325 &  0.9896 &  0.9863 &  6 &  6 \\ 
J1153+8058  & 0.8303 &  0.7782 &  0.7569 &  0.6641 &  0.6048 &  0.5600 &  \ldots &  \ldots &  0.687 &  0.6857 &  0.7557 &  \ldots &  \ldots &  32 &  6 \\ 
  & 0.8424 & 0.7940 &  0.7659 &  0.6733 &  0.6120 &  0.5691 &  \ldots &  \ldots &  0.687 &  0.6857 &  0.7557 &  \ldots &  \ldots &  6 &  \ldots \\ 
J1229+0203  & \ldots &  17.5000 &  16.9000 &  15.1000 &  14.6000 &  13.3000 &  \ldots &  \ldots &  \ldots &  \ldots &  \ldots &  10.5181 &  \ldots &  \ldots &  \ldots \\ 
J1331+3030  & 2.5063 &  2.2582 &  2.1018 &  1.7465 &  1.5331 &  1.4103 &  \ldots &  \ldots &  2.1662 &  2.1573 &  2.5561 &  1.2300 &  1.2314 &  \ldots &  \ldots \\ 
  & 2.5063 &  2.2582 &  2.1018 &  1.7465 &  1.5331 &  1.4103 &  \ldots &  \ldots &  2.1662 &  2.1573 &  2.5561 &  1.2300 &  1.2314 &  \ldots &  \ldots \\ 
J1411+5212  & 0.9449 &  0.7924 &  0.7006 &  0.5035 &  0.4119 &  0.3398 &  \ldots &  \ldots &  0.8422 &  0.8333 &  1.1159 &  \ldots &  \ldots &  \ldots &  \ldots \\ 
  & 0.9546 &  0.7868 &  0.6977 &  0.4941 &  0.3985 &  0.3324 &  \ldots &  \ldots &  0.8422 &  0.8334 &  1.1369 &  \ldots &  \ldots &  \ldots &  \ldots \\ 
J1642+6856  & 2.5780 &  2.4221 &  2.3904 &  2.0806 &  1.9259 &  1.7792 &  \ldots &  \ldots &  2.3308 &  2.3268 &  2.5172 &  1.7068 &  1.7053 &  1 &  12 \\ 
  & 2.7235 &  2.5418 &  2.5054 &  2.1874 &  1.9503 &  1.8048 &  \ldots &  \ldots &  2.3308 &  2.3268 &  2.5172 &  1.7068 &  1.7062 &  8 &  \ldots  \\ 
J1716+6836  & 0.5135 &  0.4996 &  0.5121 &  0.4865 &  0.4812 &  0.4581 &  \ldots &  \ldots &  0.5626 &  0.5630 &  0.5626 &  \ldots &  \ldots &  \ldots &  45 \\ 
J1748+7005  & 0.7770 &  0.7648 &  0.7783 &  0.7504 &  0.7474 &  0.7167 &  \ldots &  \ldots &  0.9569 &  0.9576 &  0.9569 &  \ldots &  \ldots &  3 & \ldots   \\ 
J1800+7828  & 2.4772 &  2.4091 &  2.3748 &  2.2463 &  2.1831 &  2.1160 &  \ldots &  \ldots &  1.9562 &  1.9559 &  2.0442 &  \ldots &  \ldots &  35 &  12 \\ 
  & 2.4732 &  2.4277 &  2.4050 &  2.2741 &  2.1687 &  2.0830 &  \ldots &  \ldots &  1.9562 &  1.9559 &  2.0148 &  \ldots &  \ldots &  9 &  \ldots  \\ 
J1806+6949  & 1.3699 &  1.3559 &  1.3720 &  1.3372 &  1.2859 &  1.2530 &  \ldots &  \ldots &  1.3775 &  1.3795 &  1.3775 &  \ldots &  \ldots &  3 &  3 \\ 
  & 1.3910 &  1.3600 &  1.3772 &  1.3491 &  1.2831 &  1.2318 &  \ldots &  \ldots &  1.3775 &  1.3795 &  1.3912 &  \ldots &  \ldots &  8 &  \ldots  \\ 
J1842+6809  & 0.5496 &  0.5823 &  0.6199 &  0.6549 &  0.6731 &  0.6656 &  \ldots &  \ldots &  0.5875 &  0.5889 &  0.4670 &  \ldots &  \ldots &  7 &  6 \\ 
J1849+6705  & 1.8407 &  1.8476 &  1.8987 &  1.8755 &  1.8445 &  1.8163 &  \ldots &  \ldots &  1.8643 &  1.8664 &  1.7710 &  \ldots &  \ldots &  1 &  6 \\ 
  & 1.6832 &  1.6823 &  1.7237 &  1.7140 &  1.6826 &  1.6451 &  \ldots &  \ldots &  1.8643 &  1.8664 &  1.7897 &  \ldots &  \ldots &  4 &  \ldots  \\ 
J1911$-$2007  & 1.9472 &  2.0366 &  2.1018 &  2.1430 &  2.1451 &  2.1244 &  \ldots &  \ldots &  1.9803 &  1.9836 &  1.7228 &  2.0381 &  2.0251 &  20 &  30 \\ 
J1924$-$2914  & 7.8316 &  7.4155 &  7.2218 &  6.6232 &  6.1598 &  5.9570 &  \ldots &  \ldots &  7.8556 &  7.8490 &  8.5233 &  6.0125 &  5.9952 &  27 &  27 \\ 
  & \ldots &  \ldots &  \ldots &  \ldots &  \ldots &  \ldots &  8.3900 &  6.4100 &  7.8556 &  \ldots &  8.6055 &  6.0125 &  5.9982 &  12 &  12 \\ 
J1927+6117  & 1.0695 &  1.0348 &  1.0159 &  0.9370 &  0.8856 &  0.8344 &  \ldots &  \ldots &  1.1052 &  1.1044 &  1.1604 &  \ldots &  \ldots &  4 &  7 \\ 
B1933$-$400  & \ldots &  \ldots &  \ldots &  \ldots &  \ldots &  \ldots &  1.0510 &  0.9380 &  0.9022 &  \ldots &  0.9025 &  \ldots &  \ldots &  14 &  1 \\ 
B1954$-$388  & \ldots &  \ldots &  \ldots &  \ldots &  \ldots &  \ldots &  1.5410 &  1.4580 &  1.5114 &  \ldots &  1.5277 &  1.6823 &  1.6724 &  6 &  3 \\ 
J1955+3151  & 1.2507 &  1.2283 &  1.2374 &  1.1631 &  1.1208 &  1.0563 &  \ldots &  \ldots &  1.2043 &  1.2046 &  1.2163 &  \ldots &  \ldots &  1 & \ldots   \\ 
J2000$-$1749  & 1.4553 &  1.6133 &  1.7424 &  2.0582 &  2.2130 &  2.3295 &  \ldots &  \ldots &  1.6514 &  1.6558 &  1.1549 &  2.2405 &  2.2150 &  6 &  \ldots  \\ 
  & \ldots &  \ldots &  \ldots &  \ldots &  \ldots &  \ldots &  1.4610 &  2.0730 &  1.6514 &  \ldots &  1.2828 &  2.2405 &  2.2117 &  9 &  \ldots  \\ 
J2011$-$1546  & 1.8644 &  1.7810 &  1.7287 &  1.5680 &  1.4091 &  1.3512 &  \ldots &  \ldots &  1.6371 &  1.6351 &  1.7598 &  1.4762 &  1.4742 &  11 &  2 \\ 
  &  \ldots &  \ldots &  \ldots &  \ldots &  \ldots &  \ldots &  1.9530 &  1.3900 &  1.6371 &  \ldots &  1.7598 &  1.4762 &  1.4742 &  1 &  8 \\ 
J2015+3710  & 4.4837 &  4.5687 &  4.6839 &  4.8017 &  4.8964 &  4.7478 &  \ldots &  \ldots &  \ldots &  \ldots & \ldots &  5.6103 &  5.5774 &  2  & \ldots   \\ 
J2025+3342  & 2.7372 &  2.6677 &  2.6865 &  2.6282 &  2.4944 &  2.3788 &  \ldots &  \ldots &  2.5418 &  2.5437 &  2.5799 &  2.4773 &  2.4689 &  5 & \ldots   \\ 
J2035+1056  & 0.3662 &  0.3724 &  0.3935 &  0.4195 &  0.4420 &  0.4350 &  \ldots &  \ldots &  0.4298 &  0.4308 &  0.3782 &  \ldots &  \ldots &  \ldots &  \ldots \\ 
  & 0.3721 &  0.3854 &  0.3953 &  0.4122 &  0.4408 &  0.4606 &  \ldots &  \ldots &  0.4298 &  0.4307 &  0.3868 &  \ldots &  \ldots &  \ldots &  \ldots \\ 
  & \ldots &  \ldots &  \ldots &  \ldots &  \ldots &  \ldots &  0.3970 &  0.4560 &  0.4298 &  \ldots &  0.3938 &  \ldots &  \ldots &  \ldots &  \ldots \\ 
B2047+039  & \ldots &  \ldots &  \ldots &  \ldots &  \ldots &  \ldots &  0.4900 &  0.3690 &  0.6028 &  \ldots &  0.6633 &  0.8853 &  0.8832 &  15 &  12 \\ 
J2101+0341  & 0.9552 &  0.9669 &  0.9999 &  1.0220 &  1.0348 &  1.0104 &  \ldots &  \ldots &  0.9398 &  0.9415 &  0.8646 &  1.1644 &  1.1576 &  5 &  6 \\ 
  & 0.9673 &  0.9776 &  0.9844 &  0.9809 &  0.9702 &  0.9379 &  \ldots &  \ldots &  0.9398 &  0.9407 &  0.9160 &  1.1644 &  1.1587 &  10 &  \ldots \\ 
B2106$-$413  & \ldots &  \ldots &  \ldots &  \ldots &  \ldots &  \ldots &  0.5790 &  0.3860 &  0.8273 &  \ldots &  0.9475 &  0.8923 &  0.8920 &  11 &  10 \\ 
J2107+4213  & 5.3846 &  5.2969 &  5.3540 &  5.1841 &  5.3145 &  5.0319 &  \ldots &  \ldots &  5.1875 &  5.1909 &  5.2134 &  5.5505 &  5.5207 &  \ldots &  \ldots \\ 
  & 5.4373 &  5.3228 &  5.3061 &  5.1025 &  4.8688 &  4.6650 &  \ldots &  \ldots &  5.1875 &  5.1893 &  5.3171 &  5.5505 &  5.5318 &  \ldots &  \ldots \\ 
J2123+0535  & 1.0483 &  0.9933 &  0.9814 &  0.9092 &  0.8821 &  0.8390 &  \ldots &  \ldots &  0.7480 &  0.7477 &  0.7966 &  \ldots &  \ldots &  4 &  7 \\ 
  & 1.0384 &  0.9820 &  0.9480 &  0.8631 &  0.8187 &  0.7782 &  \ldots &  \ldots &  0.7480 &  0.7470 &  0.8190 &  \ldots &  \ldots &  18 &  \ldots  \\ 
J2129$-$1538  & 0.6833 &  0.6052 &  0.5562 &  0.4366 &  0.3769 &  0.3304 &  \ldots &  \ldots &  \ldots &  \ldots &  \ldots &  \ldots &  \ldots &  \ldots &  \ldots \\ 
  & \ldots &  \ldots &  \ldots &  \ldots &  \ldots &  \ldots &  0.7000 &  0.3690 &  \ldots &  \ldots &  \ldots &  \ldots &  \ldots &  \ldots &  \ldots \\ 
J2130+0502  & 0.4797 &  0.4120 &  0.3819 &  0.2928 &  0.2391 &  0.2075 &  \ldots &  \ldots &  0.4533 &  0.4502 &  0.5646 &  \ldots &  \ldots &  2 & \ldots   \\ 
J2131$-$1207  & 1.4891 &  1.4109 &  1.3779 &  1.2916 &  1.2447 &  1.2199 &  \ldots &  \ldots &  1.6550 &  1.6541 &  1.7791 &  1.6187 &  1.6116 &  1 &  10 \\ 
  & 1.4631 &  1.3949 &  1.3654 &  1.2591 &  1.1926 &  1.1145 &  \ldots &  \ldots &  1.6550 &  1.6538 &  1.7791 &  1.6187 &  1.6149 &  15 &  \ldots  \\ 
J2134$-$0153  & 2.1876 &  2.1543 &  2.1504 &  2.0816 &  2.0294 &  1.9927 &  \ldots &  \ldots &  2.3047 &  2.3053 &  2.3507 &  1.7913 &  1.7826 &  7 &  6 \\ 
  & 2.2024 &  2.1769 &  2.1628 &  2.0700 &  1.9709 &  1.8680 &  \ldots &  \ldots &  2.3047 &  2.3048 &  2.3507 &  1.7913 &  1.7862 &  9 &  \ldots  \\ 
J2136+0041 & 5.1357 &  4.4905 &  4.3172 &  3.5545 &  3.1093 &  2.8487 &  \ldots &  \ldots &  4.4731 &  4.4569 &  5.3229 &  2.9531 &  2.9594 &  2 &  6 \\ 
  & 5.1626 &  4.6346 &  4.3453 &  3.5315 &  3.0150 &  2.6432 &  \ldots &  \ldots &  4.4731 &  4.4538 &  5.3229 &  2.9531 &  2.9623 &  21 &  \ldots  \\ 
  & \ldots &  \ldots &  \ldots &  \ldots &  \ldots &  \ldots &  5.3390 &  3.0070 &  4.4731 &  \ldots &  5.3229 &  2.9531 &  2.9579 &  26 &  24 \\ 
J2139+1423  & 1.7749 &  1.6047 &  1.5498 &  1.3154 &  1.1764 &  1.0425 &  \ldots &  \ldots &  1.6908 &  1.6863 &  1.9275 &  1.4098 &  1.4135 &  2 & \ldots   \\ 
J2146$-$1525  & 0.8719  &  0.8935 &  0.9066 &  0.9397 &  0.9524 &  0.9583 &  \ldots &  \ldots &  1.0973 &  1.0992 &  1.0314 &  1.3076 &  1.2980 &  1 &  7 \\ 
  & 0.7987 &  0.8176 &  0.8309 &  0.8472 &  0.8679 &  0.8235 &  \ldots &  \ldots &  1.0973 &  1.0988 &  1.0314 &  1.3076 &  1.3006 &  19 &  \ldots \\ 
J2148+0657  & 3.0808 & 2.9827 &  2.9578 &  2.8706 &  2.8321 &  2.8026 &  \ldots &  \ldots &  3.2672 &  3.2684 &  3.4142 &  3.1248 &  3.1080 &  1 &  \ldots \\ 
  & 3.0498 & 2.9435 &  2.9090 &  2.7485 &  2.6214 &  2.4834 &  \ldots &  \ldots &  3.2672 &  3.2667 &  3.4142 &  3.1248 &  3.1143 &  27 &  \ldots \\ 
J2151+0709  & 0.6627 &  0.6033 &  0.5712 &  0.4852 &  0.4370 &  0.4070 &  \ldots &  \ldots &  0.6933 &  0.6911 &  0.8007 &  \ldots &  \ldots &  \ldots &  \ldots  \\ 
J2158$-$1501  & 4.3713 &  4.1548 &  4.0724 &  3.7894 &  3.5564 &  3.4552 &  \ldots &  \ldots &  4.2793 &  4.2766 &  4.6002 &  3.8414 &  3.8284 &  1 &  3 \\ 
  & 4.4876 & 4.3751 &  4.2843 &  3.9684 &  3.7323 &  3.4617 &  \ldots &  \ldots &  4.2793 &  4.2766 &  4.5360 &  3.8414 &  3.8361 &  22 &  \ldots \\ 
J2248$-$3234  & 0.7983 &  0.7326 &  0.6955 &  0.6201 &  0.5706 &  0.5473 &  \ldots &  \ldots &  0.6971 &  0.6957 &  0.7946 &  1.0195 &  1.0171 &  \ldots &  \ldots \\ 
J2253+1608  & 4.6018 &  4.5424 &  4.6738 &  4.8530 &  5.0473 &  5.0780 &  \ldots &  \ldots &  5.0299 &  5.0392 &  4.9293 &  6.0161 &  5.9688 &  \ldots &  \ldots \\ 
  & 4.4691 &  4.4487 &  4.5284 &  4.7317 &  4.9101 &  4.9313 &  \ldots &  \ldots &  5.0299 &  5.0392 &  4.9293 &  6.0161 &  5.9688 &  \ldots &  \ldots \\ 
J2258$-$2758  & 3.7939 &  3.5672 &  3.5010 &  3.2700 &  3.0541 &  2.9650 &  \ldots &  \ldots &  3.6753 &  3.6733 &  3.9877 &  3.3891 &  3.3777 &  \ldots &  \ldots \\ 
  & 3.8768 &  3.7286 &  3.6316 &  3.3215 &  3.1901 &  2.9124 &  \ldots &  \ldots &  3.6753 &  3.6722 &  3.9325 &  3.3891 &  3.3811 &  \ldots &  \ldots  
\enddata
\tablecomments{Columns 1--8 present the ground-based measurements, in Jy.  In column 9, we give the Planck 28\,GHz measurements before color correction and extrapolation:  the color corrected and extrapolated values are given in columns 10 and 11.  The same pattern is followed in columns 12 and 13 for Planck 44\,GHz measurements.  Cols. 14 and 15 list the minimum intervals between Planck and ground-based measurements of a source, expressed in days, when they could easily be determined. Sources with more than a single entry were either observed twice by the VLA (late May observations listed first, early May observations below) or by both the VLA and ATCA.}
\end{deluxetable}

\end{document}

%% file: Planck.tex
\def\setsymbol#1#2{\expandafter\def\csname #1\endcsname{#2}}
\def\getsymbol#1{\csname #1\endcsname}

\def\Planck{\textit{Planck}}

\newbox\tablebox    \newdimen\tablewidth
\def\leaderfil{\leaders\hbox to 5pt{\hss.\hss}\hfil}
\def\tablenote#1 #2\par{\begingroup \parindent=0.8em
    \abovedisplayshortskip=0pt\belowdisplayshortskip=0pt
    \noindent
    $$\hss\vbox{\hsize\tablewidth \hangindent=\parindent \hangafter=1 \noindent
    \hbox to \parindent{$^#1$\hss}\strut#2\strut\par}\hss$$
    \endgroup}

\def\L2{\ifmmode L_2\else $L_2$\fi}

\def\DeltaT{\ifmmode \Delta T\else $\Delta T$\fi}
\def\deltat{\ifmmode \Delta t\else $\Delta t$\fi}
\def\fknee{\ifmmode f_{\rm knee}\else $f_{\rm knee}$\fi}
\def\Fmax{\ifmmode F_{\rm max}\else $F_{\rm max}$\fi}
\def\solar{\ifmmode{\rm M}_{\mathord\odot}\else${\rm M}_{\mathord\odot}$\fi}
\def\Msolar{\ifmmode{\rm M}_{\mathord\odot}\else${\rm M}_{\mathord\odot}$\fi}
\def\Lsolar{\ifmmode{\rm L}_{\mathord\odot}\else${\rm L}_{\mathord\odot}$\fi}
\def\inv{\ifmmode^{-1}\else$^{-1}$\fi}
\def\mo{\ifmmode^{-1}\else$^{-1}$\fi}
\def\sup#1{\ifmmode ^{\rm #1}\else $^{\rm #1}$\fi}
\def\expo#1{\ifmmode \times 10^{#1}\else $\times 10^{#1}$\fi}
\def\,{\thinspace}
\def\lsim{\mathrel{\raise .4ex\hbox{\rlap{$<$}\lower 1.2ex\hbox{$\sim$}}}}
\def\gsim{\mathrel{\raise .4ex\hbox{\rlap{$>$}\lower 1.2ex\hbox{$\sim$}}}}

\def\simprop{\mathrel{\raise .4ex\hbox{\rlap{$\propto$}\lower 1.2ex\hbox{$\sim$}}}}
\def\deg{\ifmmode^\circ\else$^\circ$\fi}
\def\pdeg{\ifmmode $\setbox0=\hbox{$^{\circ}$}\rlap{\hskip.11\wd0 .}$^{\circ}
          \else \setbox0=\hbox{$^{\circ}$}\rlap{\hskip.11\wd0 .}$^{\circ}$\fi}
\def\arcs{\ifmmode {^{\scriptstyle\prime\prime}}
          \else $^{\scriptstyle\prime\prime}$\fi}
\def\arcm{\ifmmode {^{\scriptstyle\prime}}
          \else $^{\scriptstyle\prime}$\fi}
\newdimen\sa  \newdimen\sb
\def\parcs{\sa=.07em \sb=.03em
     \ifmmode \hbox{\rlap{.}}^{\scriptstyle\prime\kern -\sb\prime}\hbox{\kern -\sa}
     \else \rlap{.}$^{\scriptstyle\prime\kern -\sb\prime}$\kern -\sa\fi}
\def\parcm{\sa=.08em \sb=.03em
     \ifmmode \hbox{\rlap{.}\kern\sa}^{\scriptstyle\prime}\hbox{\kern-\sb}
     \else \rlap{.}\kern\sa$^{\scriptstyle\prime}$\kern-\sb\fi}
\def\ra[#1 #2 #3.#4]{#1\sup{h}#2\sup{m}#3\sup{s}\llap.#4}
\def\dec[#1 #2 #3.#4]{#1\deg#2\arcm#3\arcs\llap.#4}
\def\deco[#1 #2 #3]{#1\deg#2\arcm#3\arcs}
\def\rra[#1 #2]{#1\sup{h}#2\sup{m}}

\def\dots{\relax\ifmmode \ldots\else $\ldots$\fi}
\def\WHzsr{\ifmmode $W\,Hz\mo\,sr\mo$\else W\,Hz\mo\,sr\mo\fi}
\def\mHz{\ifmmode $\,mHz$\else \,mHz\fi}
\def\GHz{\ifmmode $\,GHz$\else \,GHz\fi}
\def\mKs{\ifmmode $\,mK\,s$^{1/2}\else \,mK\,s$^{1/2}$\fi}
\def\muKs{\ifmmode \,\mu$K\,s$^{1/2}\else \,$\mu$K\,s$^{1/2}$\fi}
\def\muKRJs{\ifmmode \,\mu$K$_{\rm RJ}$\,s$^{1/2}\else \,$\mu$K$_{\rm RJ}$\,s$^{1/2}$\fi}
\def\muKHz{\ifmmode \,\mu$K\,Hz$^{-1/2}\else \,$\mu$K\,Hz$^{-1/2}$\fi}
\def\MJysr{\ifmmode \,$MJy\,sr\mo$\else \,MJy\,sr\mo\fi}
\def\MJysrmK{\ifmmode \,$MJy\,sr\mo$\,mK$_{\rm CMB}\mo\else \,MJy\,sr\mo\,mK$_{\rm CMB}\mo$\fi}
\def\microns{\ifmmode \,\mu$m$\else \,$\mu$m\fi}

\def\muK{\ifmmode \,\mu$K$\else \,$\mu$\hbox{K}\fi}
\def\microK{\ifmmode \,\mu$K$\else \,$\mu$\hbox{K}\fi}
\def\muW{\ifmmode \,\mu$W$\else \,$\mu$\hbox{W}\fi}
\def\kms{\ifmmode $\,km\,s$^{-1}\else \,km\,s$^{-1}$\fi}
\def\kmsMpc{\ifmmode $\,\kms\,Mpc\mo$\else \,\kms\,Mpc\mo\fi}
\providecommand{\sorthelp}[1]{}